\documentclass[fleqn,usenatbib]{rasti}

\usepackage[T1]{fontenc}
\usepackage[english]{babel}
\usepackage[autostyle]{csquotes}
\usepackage[normalem]{ulem}
\usepackage[version=4]{mhchem}
\usepackage{threeparttable}
\usepackage{mathptmx,lmodern,graphics,graphicx,bm,amsmath,amssymb,booktabs,dcolumn,subfig,multirow,tabularx,rotating,physics,braket,xcolor,siunitx,array,bold-extra}
\usepackage[capitalize]{cleveref}
\usepackage[format=hang,font=small,labelfont=bf]{caption}
\usepackage{svg}

\newcommand{\cm}{cm$^{-1}$}
\newcommand{\um}{$\mu$m}

\newcommand{\eg}{{\it e}.{\it g}.}

\newcommand{\abinitio}{\emph{ab initio}}
\newcommand{\Marvel}{{\sc Marvel}}
\newcommand{\Marvelised}{{\sc Marvel}ised}
\newcommand{\Marvelisation}{{\sc Marvel}isation}

\newcommand{\Duo}{{\sc Duo}}

\newcommand{\Mollist}{{\sc Mollist}}

\newcommand*{\mgo}{$^{24}$Mg$^{16}$O}
\newcommand*{\MgO}{$^{24}$Mg$^{16}$O}

\newcommand*{\mgos}{$^{24}$Mg$^{17}$O}
\newcommand*{\mgoe}{$^{24}$Mg$^{18}$O}
\newcommand*{\mgof}{$^{25}$Mg$^{16}$O}
\newcommand*{\mgoff}{$^{26}$Mg$^{16}$O}

\newcommand{\MgX}{X~$^1\Sigma^{+}$}
\newcommand{\MgA}{A~$^1\Pi$}
\newcommand{\MgB}{B~$^1\Sigma^{+}$}
\newcommand{\Mgaa}{a~$^3\Pi$}
\newcommand{\Mgb}{b~$^3\Sigma^{+}$}

\newcommand*{\VO}{$^{51}$V$^{16}$O}

\newcommand{\TiO}{$^{48}$Ti$^{16}$O}
\newcommand{\TiOisoa}{$^{46}$Ti$^{16}$O}
\newcommand{\TiOisob}{$^{47}$Ti$^{16}$O}
\newcommand{\TiOisoc}{$^{49}$Ti$^{16}$O}
\newcommand{\TiOisod}{$^{50}$Ti$^{16}$O}

\newcommand{\TiOtitle}{TiO}
\newcommand{\MgOtitle}{MgO}
\newcommand{\VOtitle}{VO}

\newcommand{\mc}{\multicolumn}
\newcolumntype{H}{>{\setbox0=\hbox\bgroup}c<{\egroup}@{}}
\newcolumntype{d}{D{.}{.}{-1}}

\graphicspath{{Figures/}}

\title[Hybrid Line Lists]{A hybrid approach to generating diatomic line lists for high resolution studies of exoplanets and other hot astronomical objects: Updates to ExoMol MgO, TiO and VO line lists}

\author[McKemmish et. al.]{Laura K. McKemmish$^{1}$\thanks{E-mail: l.mckemmish@unsw.edu.au}, Charles A. Bowesman$^{2}$, Kyriaki Kefala$^{2}$, Armando N. Perri$^{1,2}$, \newauthor{Anna-Maree Syme$^{1}$,
Sergei  N. Yurchenko$^{2}$,Jonathan Tennyson$^{2}$\thanks{E-mail: j.tennyson@ucl.ac.uk}}
\vspace*{4mm}\\
$^{1}$ School of Chemistry, University of New South Wales, 2052, Sydney, Australia\\
$^{2}$ Department of Physics and Astronomy, University College London, Gower Street, WC1E 6BT London, UK
}
\date{Accepted XXX. Received YYY; in original form ZZZ}

\pubyear{2022}

\begin{document}

\date{\today}

\maketitle

\begin{abstract}

The best molecular line lists for astrophysical applications today require both high accuracy of line positions for strong lines as well as high overall completeness. The former is required to enable, for example, molecular detection in high-resolution cross-correlation observations of exoplanets, while completeness is required for accurate spectroscopic and radiative properties over broad temperature and spectral ranges. 
The use of empirical energies generated with the \Marvel{} procedure is a standard way to improve accuracy; here we explore methods
of extending the use of these levels using predicted shifts and isotopologue extrapolation, as well augmenting the levels from other
sources such as effective Hamiltonian studies. These methods are used to update ExoMol line lists  for the main \mgo{} and \TiO{} isotopologues, as well as for \mgos{}, \mgoe{}, \mgof{}, \mgoff{}, \TiOisoa{}, \TiOisob{}, \TiOisoc{} and \TiOisod{}; new \Marvel{} results for \VO{} are also presented.


\end{abstract}


\begin{keywords}
molecular data -- techniques: spectroscopic
\end{keywords}



\section{Introduction}

The desire to characterise the atmospheres of exoplanets, and other hot astronomical objects, has led to the creation of molecular line lists (sets of energy levels and the intensity of transitions between these levels) which can be used for such studies, see the recent review by \citet{jt853}. The ExoMol project was conceived to produce line lists with high completeness at temperatures appropriate for the studies of exoplanet atmospheres \citep{jt528}; see also \citet{jt572}. Typically, such activities require spectra with a resolving power $R$ ($={\lambda}/{\Delta\lambda}$) of a few hundred rising to perhaps 3\,000 for the James Webb Space Telescope (JWST).

More recently, a new means to identify molecular species in exoplanetary atmospheres has emerged that uses ground-based high-resolution spectroscopy taking advantage of cross-correlation of observed spectra with very high-resolution template molecular spectra \citep{18Birkby,19MoSn}. This method currently provides the most secure detections of molecules in exoplanets. To be successful the cross-correlation method requires very high resolution of strong spectral lines which are known accurately enough to model observations  obtained  with $R$ values in the typical range 70\,000 to 150\,000 used for these studies. Even for diatomic species, line lists originally developed for space-based transit spectroscopy are often not suitable for these applications, as discussed by \citet{15HoDeSn.TiO} and \citet{22deKeSn.VO}.

The methodologies for constructing line lists have been adapting to meet these dual needs, crucially including explicit experimentally-derived energy levels within a high-completeness modelled line list. Details of how this is accomplished are important to enable production of the best possible data for astronomers within the limitations of current experimental data availability and accuracy of \abinitio{} electronic structure. Different methods have been developed in an ad-hoc manner to address the needs of individual molecules; it is timely now to consolidate these methodology developments and standardise terminology and nomenclature to allow comparison between different line lists. Therefore, in \cref{sct:method}, we articulate and standardise the general approach whereby the ExoMol database is updated to use the best-known energy for that quantum state from any data source (see \cref{tab:sourcetype} for common sources). This expands the now standard dihybrid \Marvel{}ised variational line list approach, which incorporates experimentally-derived energy levels obtained from the measured active vibration-rotation energy levels (\Marvel) procedure,  to a more comprehensive methodology where multiple different sources of energy levels can be incorporated into a single line list to optimise its accuracy. 

In \cref{sct:MgO,sct:TiO,sct:VO}, we demonstrate these modern approaches to line list construction for three different molecules, MgO, TiO, and VO, providing important improvements to their existing line list data, while showcasing how the approach taken differs between molecules based on existing line list qualities and spectral data availability. 
For both MgO and VO, good variational line lists existed \citep{jt759,jt644} ensuring completeness, but the accuracy of the strong line positions needed improvement; hence, the primary work in creating a better line list was in the collation of experimental data and a \Marvel{} analysis, which was subsequently used to produce a new hybrid line list.  For TiO, the heavy demand by the astronomical community, the complexity of the molecule and new experimental data warrants an update of the Toto line list  \citep{jt760} despite the original study using largely similar methodologies to the ones considered here.

\renewcommand{\arraystretch}{1.4}
\begin{table*}
    \caption{\label{tab:sourcetype}Source type abbreviations consistently applied to diatomic hybrid line lists discussed in this paper.}
    \label{tab:abbr}
    \begin{tabular}{lp{2cm}p{4cm}p{4cm}p{4cm}p{1cm}}
    \toprule
    Abbr. & Meaning & Description & Strength & Limitation & Rel. Acc.  \\
    \midrule
    \mc{3}{l}{\textit{Direct Predictions}} \\
    Ca & Calculated & Usually variational from an energy spectroscopic model, \textit{e.g.} using \Duo{} & Most extensive coverage, physically consistent extrapolation & Lower accuracy than other approaches & Low\\
    Ma & \Marvel{} & Experimentally-derived from collation of experimental assigned transitions & Highest accuracy & Limited scope, relies on experimental data availability & High \\
    EH & Effective Hamiltonian & Fitted to experimental data & Good interpolation & Lower accuracy extrapolation especially to higher vibrations & Mid\\
    Mo & MOLLIST & Usually effective Hamiltonian, but see original paper & & Method not consistent across all molecules & Mid \\
    HI & HITRAN & Mostly experimentally-derived with careful validation process, see original paper & & & High \\
    \midrule 
    \mc{3}{l}{\textit{Corrections}} \\
    PS & Predicted Shift & Ma-Ca difference (or similar) interpolated or propagated to higher rotational states within a vibronic level & Good way to correct for limitations in the spectroscopic model, \eg{} missing couplings & & Mid \\
    IE & Isotopologue-extrapolation & Ma-Ca difference (or similar) in main isotopologue propagated to other isotopologues & Accuracy obtained without additional experiments on each isotopologue & & Mid \\
    \bottomrule
\end{tabular}
\end{table*}

\section{Hybrid line list construction methodology}\label{sct:method}
\subsection{Overview} 

In this paper, we argue that the best approach for modern line list construction is an energy-hybridised line list constructed through the following procedure: 
\begin{enumerate}
    \item Produce a high-completeness line list based on a spectroscopic model fit to experimental data; within standard ExoMol format \citep{jt584,jt810} a line list should have two main components, a \texttt{.states} file which stores energy levels plus associated metadata for each state (such as degeneracy and quantum numbers) and a \texttt{.trans} file which stores Einstein A coefficients for transitions between rovibronic states. Transition wavenumbers are computed from the energy levels in the \texttt{.states} file.
    \item Update the energies of the \texttt{.states} file to provide the most accurately predicted energy of each state. A variety of data sources may be used, such as \Marvel{}
    (Measured Active Vibration-Rotational Energy Levels) \citep{jt412} experimentally-derived or effective Hamiltonian energy levels.
\end{enumerate}

The first step is well-established, with the most popular methodology involving fitting a spectroscopic model to experimental data then producing a variational line list using a nuclear motion program such as \Duo{} \citep{jt609} for diatomics, see \citet{jt511} and \citet{jt693} for further details. 

The second step is the key innovation over recent years, and the focus of this paper.  We focus on diatomic molecules here for simplicity, but the methodology is general for rovibronic and rovibrational line lists for small molecules. 

We use the term ``source type'' to reference the source of the energy level prediction. There are two main classes of predicted energy levels, variational calculations and empirical such as from \Marvel, plus corrections that directly improve predicted energy levels by taking note of observed uncertainties between high-accuracy and low-accuracy direct predictions. Though different terminology has been used in the literature, we standardise this in \cref{tab:sourcetype} and in particular have updated line lists on the ExoMol website (\url{www.exomol.com}) to provide consistent two-letter abbreviations for the source of energy levels within the \texttt{.states} files. 

Note that, regardless of the source type for the energy level, transition intensities are calculated using dipole moment curves that together with a particular model for the energy levels allow the Einstein A coefficient of each transition to be calculated. Dipole moment curves are almost exclusively derived from \abinitio{} calculations, which have been shown to be highly accurate in favourable cases \citep{jt613,jt871} and their use compensates for unreliability or unavailability of absolute intensity measurements
for most of the molecules considered in the ExoMol database. However, in some cases measurements of excited state lifetimes can be used to scale these \abinitio{} dipoles
\citep{jt921,jt923}.



\subsection{Source types for energy levels}\label{subsec:source_types}

In this section, we detail the source types for energy levels included in the three line lists updated in this paper: MgO, TiO and VO. Of particular importance is the consideration of uncertainties. 

\subsubsection{Experiment energy levels (often collated by \Marvel{} analysis)}

\paragraph*{Description}
The experimental methodology relies on measurements obtained astronomically or in the laboratory. By the late twentieth century, this approach led to an abundance of experimental studies which were largely left as independent sources of data, which are extremely challenging for astronomers to utilise. 

\citet{jt412} developed the \Marvel{} procedure to facilitate the provision of empirical energy levels with experimental accuracy. This procedure involves the collection of assigned, high-resolution experimental transition frequencies from all experimental studies in the literature. Using the \Marvel{} procedure, the full breadth of experimental studies can be successfully collated into a self-consistent spectroscopic network of energies and transition frequencies with reliable uncertainties.  The \Marvel{} algorithm  inverts measured frequencies, with specified uncertainties, to give the empirical energy levels of the molecule. This is achieved using a weighted least-squares fit and produces a self-consistent spectroscopic network that accounts for all observed transitions \citep{11CsFuxx.marvel}. The \Marvel{} algorithm employs graph theory \citep{00AlHaS1.TiO} and thus makes no model assumptions but relies exclusively on the rovibronic transition labels provided. 

\paragraph*{Discussion}

The experimental methodology almost certainly produces the most accurate spectral frequencies data for a molecule where available (though absolute intensities are far less reliable).

However, the time and cost associated with producing experimental spectral data is very high and the number of lines measured tends to be very small compared to the full number of transitions that are important for describing molecular opacity, particularly at higher temperatures. 

Use of the \Marvel{} procedure to obtain high accuracy energy levels can enable a small number of transitions to be used to  predict the transition frequencies of a much larger number of transitions while retaining the low experimental uncertainties.

\paragraph*{Utilisation} 
The \Marvel{} procedure has proved indispensable in producing a single comprehensive experimental line list from the plethora of experimental data available in the literature. There are currently 18 diatomic molecules of astronomical importance for which \Marvel{} datasets are available: C$_2$ \citep{jt637,jt809}, TiO \citep{jt672}, AlH \citep{jt732,jt922}, SiO \citep{jt847}, AlO \citep{jt835}, NO \citep{jt868,jt831}, BeH \citep{jt722}, O$_2$ \citep{19FuHoKoSo}, CN \citep{20SyMc}, NH \citep{jt764}, ZrO \citep{jt740}, VO \citep{jt869}, CH \citep{jt868}, OH \citep{jt868}, SiN \citep{jt875}, SO \citep{jt924}, YO \citep{jt921} and CH$^+$ \citep{jt913}.


\paragraph*{Uncertainties}

A central component of the \Marvel{} procedure is the propagation of errors from the measured transition frequencies to errors in the empirical energy levels. This means that the \Marvel{} procedure automatically determines the uncertainty of energy levels based on the input uncertainties of transitions. Self-consistency in the spectroscopic network is forced through the validation process in \Marvel{}, helping to ensure that unrealistically low transition uncertainties are corrected. 

The method of predicting uncertainties can change between \Marvel{} versions. In this work we employed the latest version of \Marvel{}, {\sc Marvel4.1}, which uses a bootstrap method to determine the uncertainties
associated with each final energy level \citep{jt908}.

For most molecules, \Marvel{} uncertainties have been adopted in ExoMol line lists without modification as an estimate of the uncertainty of that energy level. This approach makes the implicit assumption that the uncertainties of the initial ($\Delta \tilde{E}_i$) and final ($\Delta \tilde{E}_f$) state energy value can be used to give the uncertainty of the observable transition frequency ($\Delta \tilde{\nu}$) as
\begin{equation}
    \Delta\tilde{\nu} = \sqrt{(\Delta \tilde{E}_i)^2 + (\Delta\tilde{E}_f)^2}.
\end{equation}

\subsubsection{Calculated variational energy levels (best compiled by \Duo{} for coupled electronic states)}

\paragraph*{Description}
The variational  methodology of line list construction relies on the numerical solution of the nuclear motion Schr\"{o}dinger equation. This requires a preliminary solution to the electronic structure problem which yields an electronic spectroscopic model comprising a set of potential energy, coupling and dipole moment curves that vary with the geometric parameters of the molecule. For diatomic molecules, this model is simply a function of bond length. Although the energy spectroscopic model can be obtained purely from \abinitio{} electronic structure calculations, the resulting \abinitio{} variational line list often deviates significantly from an experimental line list. This is especially true for open-shell electronic states \citep{jt632}. Therefore, it is essential to replace \abinitio{} curves with empirical curves that are fitted to experimental data whenever possible. This fitting procedure can be a lengthy process that requires many iterations and some spectroscopic intuition.

\paragraph*{Discussion}
The variational methodology provides the most complete line list that extrapolates well to high energy. However, even after fitting to experimental data, variational line lists do not typically achieve spectroscopic accuracy and are typically less accurate than perturbative methods in regions where effective Hamiltonians perform well.

\paragraph*{Utilisation}
The ExoMol database \citep{jt528,jt810} is the most comprehensive source of variational spectroscopic data. This database contains spectroscopic models and variational line lists for over 60 diatomic molecules. These data are largely generated using the \Duo{} variational nuclear motion program \citep{jt609}, although LEVEL \citep{LeRoy2017LEVEL:Levels} has also been used for some (largely closed shell) molecules.

\paragraph*{Uncertainties}
Uncertainties for variational line list data are very dependent on the molecule and energy level. Typically, uncertainties are best evaluated by comparing variational data with experimentally-derived energy levels (like from \Marvel{}) whenever possible and then extrapolating to similar rovibronic states with no available experimental data, increasing uncertainties with $v$ and $J$.

Electronic states for which there is no available experimental data have the largest uncertainty as this relies on the cancellation of errors in the quantum mechanical treatment between states usually of different spin symmetry. For 3d transition metal diatomics with modern quantum chemistry approaches, the review by \citet{jt632} found that errors in excess of 1000 \cm{} are common and thus purely computational predictions are clearly unsuitable for direct comparisons with experimental spectra. Typically, these states are included to increase the accuracy of the partition functions (which are not as sensitive to errors) and to understand the scale of the missing absorption to direct future experiments.

The uncertainty in the position of excited vibrational states purely known from experiment is smaller, with fundamental frequencies typically predicted to an accuracy of around 20 \cm{} or better theoretically \citep{jt632} even for complex 3d transition metals and much better for smaller molecules. Nevertheless, this is often not accurate enough to model absorption bands at the resolution needed for JWST and thus having experimental data at least for the $v=0,1$ and ideally the $v=0,1,2$ levels is a strong advantage. 
We note that, for well-studied molecules, typically the available experimental data is for the strongest absorption bands as these are the easiest experiments.

Uncertainties are estimated for variationally calculated levels assuming that the errors grow linearly
with vibrational excitation but quadratically with rotational excitation; these assumptions accord with our previous
experience for a variety of this species, as first adopted in \citet{jt804}. We therefore estimated uncertainties using the expression:
\begin{equation}
    \label{eq:unc_ca}
    \Delta \tilde{E}_{\text{Ca}} = a \cdot v + b \cdot J(J+1) + c,
\end{equation}
where $a$, $b$ and $c$ are constants in units of \cm{}. $c$ is the initial minimum uncertainty determined for each electronic state. $a$ and $b$ quantify the uncertainty scales for the vibrational $v$ and rotational $J(J+1)$ parameters, respectively.

\subsubsection{Predicted shift correction} \label{sec:PS}

\paragraph*{Description}

Levels in variational line lists are generally computed to a higher maximum total angular momentum quantum number $J$ than are
available from experimental data.
While \textit{ab initio} calculations and refinement to experimental energies often yield highly accurate energies, extrapolations must still be made to levels outside the observed range.
It is also often the case that experiments do not provide complete level coverage up to the highest observed $J$ level.
An improved estimate of the energies of the unobserved rotational states can be obtained by applying a shift to the calculated energy levels based on the trends in the observed minus calculated (obs.-calc.) energy differences in that band, where ``obs.'' refers to the experimentally-derived \Marvel{} energy levels and ``calc.'' to the \Duo{} variationally computed energy levels.
This is done by fitting to the obs.-calc. trends in a given spin-orbit and parity component of a vibronic state, defined by the electronic state, vibrational quantum number $v$, $\Omega$ and rotationless parity, and predicting synthetic obs.-calc. values for the missing levels.

\paragraph*{Discussion}

Applying these synthetic energy shifts to the variationally calculated energies allows for a more extensive provision of high-accuracy energy levels, particularly when interpolating energy shifts for levels within the $J$ range of experimental data.
With this method, conservative energy shifts can be predicted for levels outside the observed $J$ range, extrapolating to the maximum $J$ in the variational calculation.
This also allows for predictions to be made for levels with lower values of $J$ than are observed in experiment, which is particularly useful in cases where low-$J$ transitions lie in clustered band heads and cannot be assigned.
Similarly, this approach is effective at correcting the calculated energies of levels near electronic perturbations and resonances, which can often be difficult to accurately model in a spectroscopic model.
Applying this method removes any potential discontinuities between \Marvelised{} and calculated energy levels, which is particularly important for states that have a poor fit to experimental data.

\paragraph*{Utilisation}
The predicted shift methodology has been employed in the construction of high-resolution line lists for AlO \citep{jt835} and VO \citep{jt923}.
This AlO line list has been successfully utilised in the detection of near-infrared electronic bands in the atmosphere in an eruptive young stellar object \citep{jt928}.

\paragraph*{Uncertainties}

The uncertainties for PS levels for every state within the $J$ range for which we have experimental data were calculated the same way as the energy values (see \cref{sec:PS}), as a fit of the \Marvel{} uncertainties.
When extrapolating to lower $J$ values, the PS level uncertainties were estimated to be simply equal to the standard deviation $\sigma$ of the obs.-calc.

Outside this range, the PS energy uncertainties were estimated as a function of the standard deviation of the known obs.-calc. data point in the band, $\sigma$.
This $\sigma$ value was used as a starting minimum uncertainty for the extrapolation; uncertainties were scaled for a level based on how much larger its $J$ value was than the maximum $J$ that occurred for the \Marvel{} data in that band, $J^\text{{Ma}}_\text{{max}}$.
Taking $J_{\text{ext}} = J-J^\text{{Ma}}_\text{{max}}$, the following equation was used:
\begin{equation}\label{e:ps_unc} 
    \Delta \tilde{E}_{\text{PS}} = a \cdot J_{\text{ext}}(J_{\text{ext}}+1) + \sigma,
\end{equation}
where $a$ and $\tilde{E}_{\text{PS}}$ are determined for each individual molecule.

\subsubsection{Isotopologue-extrapolation correction}\label{s:phm}

\paragraph*{Description}
A spectroscopic model is often only optimised for the main isotopologue of a given molecule due to the lack of assigned experimental spectral data for other isotopologues. Computationally, however, a variational line list for the other isotopologues can easily be produced by changing the mass(es) of the atoms in the main spectroscopic model. Consequently, an improved pseudo-hybrid line list for an isotopologue can be generated by shifting each energy with reference to the corresponding energies in the main variational and hybridised line lists  \citep{jt665} using
\begin{equation}\label{e:iso}
    \tilde{E}^\text{iso}_\text{IE} = \tilde{E}^\text{iso}_\text{Ca} + \left(\tilde{E}^\text{main}_\text{hybrid} - \tilde{E}^\text{main}_\text{Ca}\right),
\end{equation}
where $\tilde{E}^\text{main}_\text{Ca}$ is the calculated variational energy of the main isotopologue, $\tilde{E}^\text{main}_\text{hybrid}$ is its hybridised energy, $\tilde{E}^\text{iso}_\text{Ca}$ is the calculated variational energy of a given isotopologue and $\tilde{E}^\text{iso}_\text{IE}$ is its isotopologue-extrapolated (IE) energy. In other words, it is assumed that the energy residuals between the hybridised and variational line lists are constant for all isotopologues.

\paragraph*{Discussion}
Isotopologue-extrapolation is a very effective way to provide improved accuracy for isotopologue spectra, relying on the fact that the impact of nuclear mass on vibrational energy levels is well understood and that the underlying potential energy surface is to a high level of approximation identical.

\paragraph*{Utilisation}
The usefulness of the isotopologue methodology can perhaps best be illustrated by considering TiO, for which the dominant isotopologue typically only comprises 73\% of abundance but is the subject of almost all experimental studies. By using the isotopologue-extrapolated method, line lists for \TiOisoa{}, \TiOisob{}, \TiOisoc{} and \TiOisod{} \citep{jt760} were successfully utilised by \citet{jt799} to obtain Ti isotopic abundances in a number of spectral regions for two M-dwarf stars. They have also been employed in the successful detection of TiO in the atmospheres of several hot Jupiter exoplanets 
\citep{20EdChBa.exo,21SeNuMo.exo,22BiHoKi.exo,24EdChxx.exo}.

Note that the isotopologue-extrapolation method has been in use since at least \citet{jt665}, who describe the procedure
as pseudo-experimental, with various terminology. This paper intends to standardise this for clear communication \citep{21McKemmishDiatomics}.

\paragraph*{Uncertainties}
The uncertainties for IE levels are typically scaled by a constant with respect to those of the main isotopologue. This constant may be determined through comparison to experimental data if available. 

For Ca uncertainties, \cref{eq:unc_ca} should be applied for each isotopologue with modified $a$, $b$ and $c$ parameters. This is because isotopologues with a reduced mass greater than that of the main isotopologue will in general have a higher density of states and it thus may not always be possible to match all energy levels by assignment.






\subsection{Construction of hybrid line lists} 

\subsubsection{Approach}

Our goal is to produce a line list that is as complete and as accurate as possible, which we achieve by combining multiple sources of data. Specifically, the aforementioned experimental, perturbative and variational line lists are collated into a hybrid line list with energy levels then modified using predicted shifts; isotopologue line list accuracy can be enhanced by using the isotopologue-extrapolation method described above. The hybrid~line lists produced in this way exploit the unique advantages of each methodology and ameliorate the limitations in order to produce the most~accurate and comprehensive source of spectroscopic data for a given molecule. 



To construct the final \texttt{.states} file for a hybridised line list, we start with a set of \texttt{.states} files in a standard and consistent ExoMol format \citep{jt584,jt810}, each containing predicted energies with uncertainty estimates for all quantum states of interest. The length of each of these \texttt{.states} files is expected to vary with the highest-quality data (usually \Marvel) typically having the fewest energy levels and providing the desired high-accuracy spectral information, while the lowest-quality data (usually variational) incorporates all energy levels considered within the line list and provides the completeness for the final line list. Using a loop over all quantum states, a final hybridised \texttt{.states} file is constructed by using the highest-quality energy level prediction available for each state and assigning the appropriate label and predicted uncertainties. The highest-quality data can be identified either by comparison of the compiled uncertainties or through a pre-determined hierarchy (\eg{} experimental data is more accurate that perturbative data which is more accurate than variational data).

\subsubsection{Challenges with hybridisation}

At this stage, significant deviations of the predicted energies from different data sources for the same quantum state should be flagged and manually reviewed. Misassignments of experimental transitions can be uncovered. However, more commonly, there is a misalignment between the sources in determining the quantum numbers used to describe each rovibronic state. Both issues are more common in complex diatomic molecules where there is strong mixing between electronic states. Anecdotally, differences in assignment of quantum numbers between \Duo{} computationally-assigned variational and model Hamiltonian assigned experimental states becomes quite common for large $J$ especially when electronic states are coupled like for transition metal diatomic molecules with complex electronic structures. On occasion it has proved necessary to perform a thorough reassignment of the experimental transitions to obtain a consistent solution, see \citet{jt618} for example.

The origin of this challenge is that most of the quantum numbers used to describe states are not rigorous and thus must be defined approximately manually or by computer programs such as \Duo{}. Formally, as \Duo{} diagonalises the total rovibronic Hamiltonian using Hund’s case (a) functions, a rovibrational state with a given electronic state can be uniquely defined (in the absence of hyperfine couplings) using its total parity, vibrational $v$ quantum number, total angular momentum excluding nuclear spin $J$ quantum number and its projection along the internuclear axis $\Omega$ ($=\Lambda+\Sigma$) quantum number. Aside from the rigorously defined $J$ and parity, $v$ and $\Omega$ are simply convenient labels that are assigned based on dominant contribution to the total wave function. From experience, the misassignment of these two quantum numbers becomes common for sufficiently large $J$, especially in cases where mixing arises from large spin-orbit or other couplings. Due to the overall completeness of the variational methodology, however, the rovibronic states population can be trusted (i.e. all states are calculated without any gaps), although the energy of each state may not be consistent with experimental data. For this reason, $v$ and $\Omega$ may be simply reassigned based on energy orderings.


\begin{table*}
    \caption{A part of the final \texttt{.states} file for \MgO. The same structure is used for TiO and VO line lists. }
    \label{tab:states}
    \begin{threeparttable}
    \begin{center}
    \resizebox{\linewidth}{!}{
    \begin{tabular}{rrrrrrrrrlccccrr}
    \hline
    $i$ & $\tilde{E}$ & $g_\text{tot}$ & $J$ & unc & $\tau$ & $g$ & $p_{+/-}$ & $p_{e/f}$ & State & $v$ & $\Lambda$ & $ \Sigma$ & $\Omega$ & Label & $\tilde{E}_{\text{Ca}}$ \\
    \hline
    1 & 0.000000 & 1 & 0.0 & 0.000000 & NaN & 0.000000 & + & e & X(1SIGMA+) & 0 & 0 & 0.0 & 0.0 & Ma & 0.000000\\
    2 &	774.738739 & 1 & 0.0 &	0.001420 & 1.3596E+00 &	0.000000 &	+	& e &	X(1SIGMA+) &	1 &	0 & 0.0 & 0.0 & Ma & 774.737704\\
    3 & 1539.174552 & 1 & 0.0 &	0.001730 &	6.3331E-01 & 0.000000 & + &	e &	X(1SIGMA+) & 2 &	0 &	0.0 & 0.0 & Ma & 1539.136871\\
    4 & 2292.687940	& 1	& 0.0	& 0.003610 & 3.9854E-01	& 0.000000	& +	& e	& X(1SIGMA+) &	3	& 0 &	0.0	& 0.0 & Ma & 2293.035126\\
    5 &	2621.208578	& 1	& 0.0 & 0.004809 &	2.5844E-02 & 0.000000	& + &	e &	a(3PI) &	0	& 1	& -1.0 & 0.0 & PS & 2621.207376\\
    6 &	3037.070194 &	1 &	0.0 & 0.040000 & 2.2228E-01	&  0.000000	& +	& e	& X(1SIGMA+) &	4 &	0	& 0.0	& 0.0 & Ca & 3037.070194\\
    7 &	3265.064334	& 1	&0.0	& 0.043377 &	2.0867E-02 & 0.000000 & + & e & a(3PI) & 1 &	1 & -1.0 &  0.0 & PS & 3264.977473\\
    8 &  3770.264143 &	1 & 0.0 & 0.050000 &  9.7014E-02 &	0.000000	& + & e	& X(1SIGMA+) &	5 &	0 &	0.0 &	0.0 & Ca & 3770.264143\\
    9 &	3900.848866 & 1 & 0.0 & 0.030000 &	1.8289E-02 & 0.000000 & +	& e	& a(3PI) & 2 & 1	& -1.0	& 0.0 & Ca & 3900.848866\\
    10 & 4478.700487	& 1	& 0.0	& 0.040000    &	2.3189E-02 & 0.000000 & +	& e &	a(3PI) &	3 &	1	& -1.0 & 0.0 & Ca & 4478.700487\\
    11 &	4543.125919	& 1	& 0.0	& 0.060000    & 2.7738E-02 & 0.000000	& +	& e	& X(1SIGMA+)	&6	&0	&0.0	&0.0 & Ca & 4543.125919\\
    \hline
    \end{tabular}
    }
    \begin{tablenotes}
      \item [a] $i$: State counting number
      \item [b] $\tilde{E}$: Energy (in \cm)
      \item [c] $g_\text{tot}$: Total degeneracy
      \item [d] $J$: Total angular momentum
      \item [e] unc ($\Delta \tilde{E}$): Uncertainty (in \cm{})
      \item [f] $\tau$: Lifetime (in s; calculated in \textsc{ExoCross} \citep{jt708})
      \item [g] $g$: Lande g-factor
      \item [h] $p_{+/-}$: Total parity
      \item [i] $p_{e/f}$: Kronig rotationless parity
      \item [j] State: Electronic state
      \item [k] $v$: Vibrational quantum number
      \item [l] $\Lambda$: Projection of electronic orbital angular momentum on the internuclear axis
      \item [m] $\Sigma$: Projection of electronic spin angular momentum on the internuclear axis
      \item [n] $\Omega$: Projection of the total angular momentum excluding nuclear spin along the internuclear axis
      \item [o] Label (see \cref{tab:abbr}): Ma for experimental \Marvel{} energies, PS for energies from Predicted Shift method, Ca for unchanged calculated\\ energy by \cite{jt759}
      \item [p] $\tilde{E}_{\text{Ca}}$: \Duo{} calculated energy by \cite{jt759}
    \end{tablenotes}
    \end{center}
    \end{threeparttable}
\end{table*}

\subsubsection{Line list data product}

At the conclusion of the energy-hybridised line list construction process, a final \texttt{.states} file is constructed that typically looks similar to \cref{tab:states} (extract from the \MgO{} line list constructed in Section 3 of this paper). 

As stated above, the final hybrid line list retains the \texttt{.trans} file from the variational line list.

Each ExoMol line list is given a name and version number (the date in YYYYMMDD format) which appears in the definitions (\texttt{.def}) file for each isotopologue/line list combination \citep{jt939}. By convention updates to the \texttt{.states} file retain the line list name but generate a new version number. Any changes to \texttt{.trans} file, in particular to the Einstein A coefficients, would result in a new name for the line list. Due to energy changes arising from the hybridisation procedure, transitions may yield negative frequencies after updating an existing \texttt{.states} file. These transitions have been removed herein without changing the line list name.











\subsection{Analysis: Suitability of line list for molecule detection using  high-resolution cross-correlation techniques}
\label{sec:twofigs}
In order to ensure effective utilisation of ground-based telescopes, astronomers need to know where each line list can be trusted for high-resolution studies. Similarly to support these efforts, experimentalists need to know the most important gaps in existing data to target to improve coverage and accuracy. 

To support these goals, here we describe two relatively new powerful visualisations and how these should be interpreted; these visualisations have already been produced for some recent molecular line lists including ZrO \citep{perri2023full} and NH \citep{perri2024full}. 

Both visualisations aim to understand the suitability of a line list for high-resolution cross-correlation studies (HRCC) by looking at the transitions whose frequencies are predicted to high accuracies. The simplest approach, which we adopt in this paper, is to consider all \Marvel{} energy levels to be sufficiently accurate for HRCC studies and also consider the predicted shift energy levels to be potentially sufficiently accurate. This is only an approximation, however, as some experimental energy levels might have higher uncertainties; for example, for MgO, only $21.4\%$ of the transitions between the \Marvel{} energy levels have a resolving power $R = \lambda/\Delta\lambda$ larger than $100\,000$. 

\subsubsection{Transition source type plots}
The transition source type plots shown in \cref{fig:mgo_cumulative,fig:tio_cumulative,fig:vo_cumulative} show cumulative density of transitions as a function of intensity figures are shown in . These plots should be read by looking at vertical slices - these visually compare the data sources used to predict the frequencies of all transitions with intensity larger than the x axis value. For example, taking a vertical slice at x = 10$^{-20}$ cm/molecule in the TiO figure, we are considering \textit{all} transitions in \TiO{} with intensities \textit{above} $10^{-20}$ cm/molecule at 2000 K (i.e. not just those transitions binned in this intensity range). The colours and y axis shows the origin of the energy of the upper and lower quantum state for these transitions; e.g. 20\% of the transitions are "Ma-Ma" meaning the upper and lower state energies are both obtained from a \Marvel{} analysis and thus highly reliable, while approximately 30\% are "Ma-Ca" meaning one energy level is from \Marvel{} and one is from Duo.

In the transition source type plots, the frequencies for transitions with source "Ma-Ma" are highly accurate (and thus suitable for molecule detection using high-resolution cross-correlation techniques), while other transitions with other sources have decreasing accuracy in frequency down the legend (e.g. "Ca-Ca" have least accurate transition frequencies).

Typically, the strongest transitions have far more highly accurate "Ma-Ma" transitions than weaker transitions. This occurs because experimentally more intense transitions are easier to observe. Fortunately, strong transitions are also the transitions for which high accuracy in frequencies is most important in cross-correlation analyses.

\subsubsection{High-res vs total cross-section plots}
The second type of plot that is extremely useful for the analysis of a line list's suitability for high-resolution cross-correlation (HRCC) is the high-resolution (high-res) vs total cross section plots, shown in \cref{fig:mgo_decomp_ma_ps,fig:tio_decomp_ma,fig:vo_decomp_ma_ps}. The dark black line shows the total cross-section for the line list while the coloured line shows the cross-section as a function of wavenumber (or equivalently wavelength on the upper horizontal axis) obtained only using transitions with frequencies known to very high precision (usually \Marvel{} energies for both the upper and lower state). For the temperature considered, in those spectral regions where the coloured and black line visually overlap, the current line list is entirely suitable for HRCC  analysis; in contrast regions where the coloured line is much lower in intensity than the black line are not suitable for cross-correlation analysis because many intense transitions are not known to very high experimental accuracy. 

It is important to note that these high-res vs total cross-section plots will change substantially at different temperatures. However, these plots can be produced at a given temperature using ExoCross using filtering based on the source type of the upper and lower energy levels.

\section{\MgOtitle{}: Increasing accuracy by adding \Marvel{} energy levels}\label{sct:MgO}

\subsection{Background}

Due to the cosmic abundance of magnesium and oxygen and the strength of the bond, diatomic magnesium oxide (MgO) in the gas phase is known to be present in the upper atmosphere of Earth \citep{08CoAiGr.MgO} and Mercury \citep{11SaKiMc.MgO,12stmcti.mgo}, where it is believed to be produced by micrometeoroid impact. However, searches for gaseous MgO in the interstellar medium have thus far been unsuccessful \citep{85TuStxx.MgO,98SaWhKa.MgO}.

In its solid form, magnesium oxide is thought to be one of the most abundant rocks in the interiors of planets \citep{23PaHuVa.MgO} and to be a component of interstellar dust \citep{82MaDuxx.MgO,15Moajal.MgO, 95YoGrxx.MgO, 03NoKoUm.MgO, 09Rietmeijer.MgO}; certainly it is an important part of Earth's mantle \citep{08CoAiGr.MgO}, but the chemical identity of solid species is difficult to confirm remotely.

However, in ultra hot rocky exoplanets like hot rocky super-Earths, often referred to as lava or magma worlds, the higher temperatures in the atmosphere are thought to favour the production of MgO vapour \citep{jt693,12ScLoFe.MgO, 22ZiVaMi.exo, 24FaTrCh.MgO}. With high-quality spectral line lists, these hypotheses can be experimentally tested. 

\citet{jt759} produced the LiTY variational line list covering the isotopologues \mgo,  \mgos, \mgoe, \mgof\ and \mgoff.   While this line list was heavily tuned to experimental data, no formal \Marvel{} project was undertaken. This line list thus has the completeness and accuracy necessary for molecular detection in moderate resolution observations, but not the accuracy needed for high-resolution cross-correlation studies. 

The existing LiTY spectral data is likely already sufficient for enabling detection with space-based telescopes like JWST. However, more definitive detection is obtained with ground-based telescopes using high-resolution cross-correlation techniques; this is enabled by the updates of this paper to the LiTY line list data specifically the explicit inclusion of \Marvel{} energies.

Here, we facilitate the detection of MgO in high-resolution studies by performing a \Marvel{} analysis of \MgO{} and updating the line list to get spectroscopic accuracy for many of the strong spectral lines.

\subsection{Data sources for energy levels}

\subsubsection{Experimental energy levels from a new MgO \Marvel{} analysis}

\begin{table*}
    \caption{Overview of the MgO \Marvel{} compilation energy levels (EL) and comparison against \Duo{} variational calculated energy levels (|Ma-Ca|), in \cm.}
    \label{tab:MgOMarvel}
    \begin{tabular}{lccrccccccccc}
        \toprule
        State & $v$ range & $J$ range & \#ELs & Unc. range & Avg. Unc. &  EL range & |Ma-Ca| range & Mean |Ma-Ca| \\ 
        \midrule
        \MgX & 0-4 & 0-61 & 125 & 0-0.25 & 0.06 & 0-3042.13 & 0-1.54 & 0.07 \\
        \MgA & 0-3 & 1-62 & 368 & 0.001-0.25 & 0.09 & 3504.32-7055.19 & 0.0003-0.1 & 0.02 \\
        \MgB & 0-1 & 0-42 & 71 & 0.003-0.14 & 0.06 & 20003.59-21387.62 & 0.00004-0.06 & 0.02 \\
        \Mgaa & 0-1 & 1-43 & 256 & 0.001-0.15 & 0.07 & 2551.29-4129.27 & 0.00006-0.33 & 0.03 \\
        \bottomrule
    \end{tabular}
\end{table*}

\begin{figure}
    \centering
    \includegraphics[width=\linewidth]{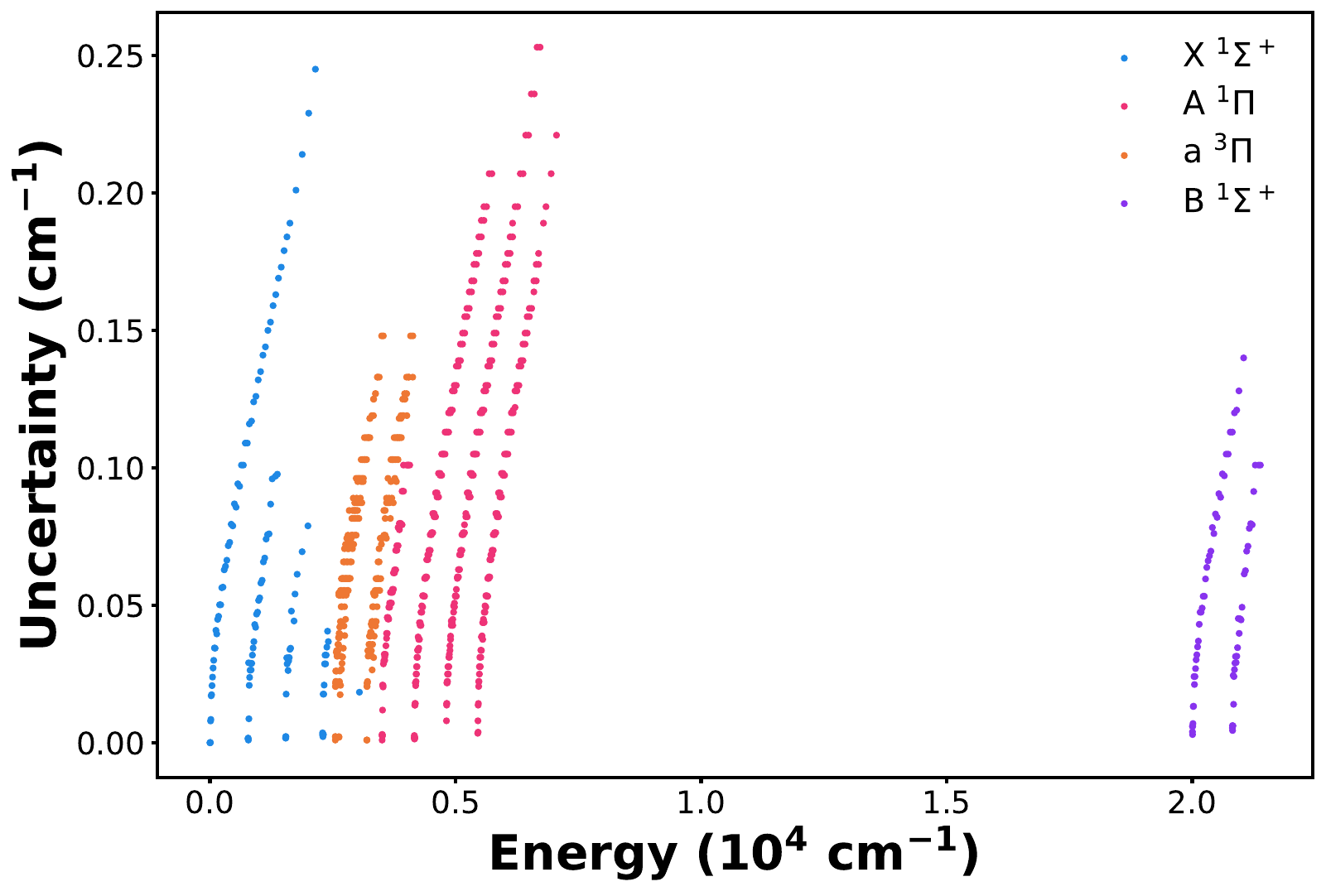}
    \caption{The \Marvel{} (Ma) energy uncertainties for \MgO{} as a function of energy.}
    \label{fig:MgO_unc}
\end{figure}

A \Marvel{} analysis for the spectroscopic data of \mgo\  was performed with results summarised in \cref{tab:MgOMarvel}. We used {\sc Marvel4.1} and produced the \Marvel{} energy uncertainties with the bootstrapping method as described in \citet{jt908}. The \Marvel{} uncertainties are shown in \cref{fig:MgO_unc}.

A total of 1181 transitions were collected from 9 sources \citep{84AzDyGe.MgO,84StAzCa.MgO,86ToHoxx.MgO,91IpCrFi.MgO,91CiHeBl.MgO,94KaHiTa.MgO,94MuRiPf.MgO,95MuThPf.MgO,06KaKaxx.MgO}. 1169 of these transitions are validated through the \Marvel{} procedure and inverted to determine 820 empirical energy levels and uncertainties for the four lowest electronic states (\MgX, \Mgaa, \MgA, \MgB) and  vibrational levels $v=0-4$, with the highest rotational quantum number being $J_{\rm max}=62$. 

The relevant details can be found within the supporting information, including a full list of all the experimental data used in this analysis, a segment file connecting the sources to the wavenumber units used, a justification of the uncertainties considered for the line positions, a list of our \Marvel{} energy levels, and a list of papers considered but not used in the current work as well as the reasons for their exclusion. 

For energy levels with \Marvel{} replacement, uncertainties were taken from the \Marvel{} procedure (see \cref{fig:MgO_unc}).

\subsubsection{Variational energy levels}

Here we retain the spectroscopic model used for the LiTY line list for \MgO{} in \citet{jt759}. 
However, the availability of high-quality experimental data allows an assessment of the accuracy of the spectroscopic model and thus a improves the quality of our predicted uncertainties. 
For the four lowest-lying electronic states we find a high accuracy, with an average error of 0.029 \cm{} and maximum value 0.2 \cm{} within the states for which we had experimental data and applied the PS method (see \cref{tab:MgOMarvel} for $J$ and $v$ ranges). For all other cases, the errors exceed this value. For calculated levels, the errors are within 2 \cm{} for $J$ values up to 140 and increase to a maximum of 10 \cm{} for the highest $J$ values ($J > 300$).

The uncertainties of the calculated levels (Ca) were considered to have a linear dependence on $v$ and a quadratic dependence on $J$ following the relation given in \cref{eq:unc_ca}.
Here, we adopted values of $a = 0.0001$ \cm{} and $b = 0.01$ \cm{} for all electronic states, $c = 0$ \cm{} for the ground state, and $c = 0.01$ \cm{} for all excited electronic states to estimate the uncertainties, with a maximum cutoff of 10 \cm.
Unfortunately, due to a lack of experimental data for the \Mgb{} state we have little information to base the uncertainties of the energy levels in this state on; we estimated an uncertainty of 10 \cm{} for all energy levels in this state.

\begin{figure*}
    \centering
    \includegraphics[width=\textwidth]{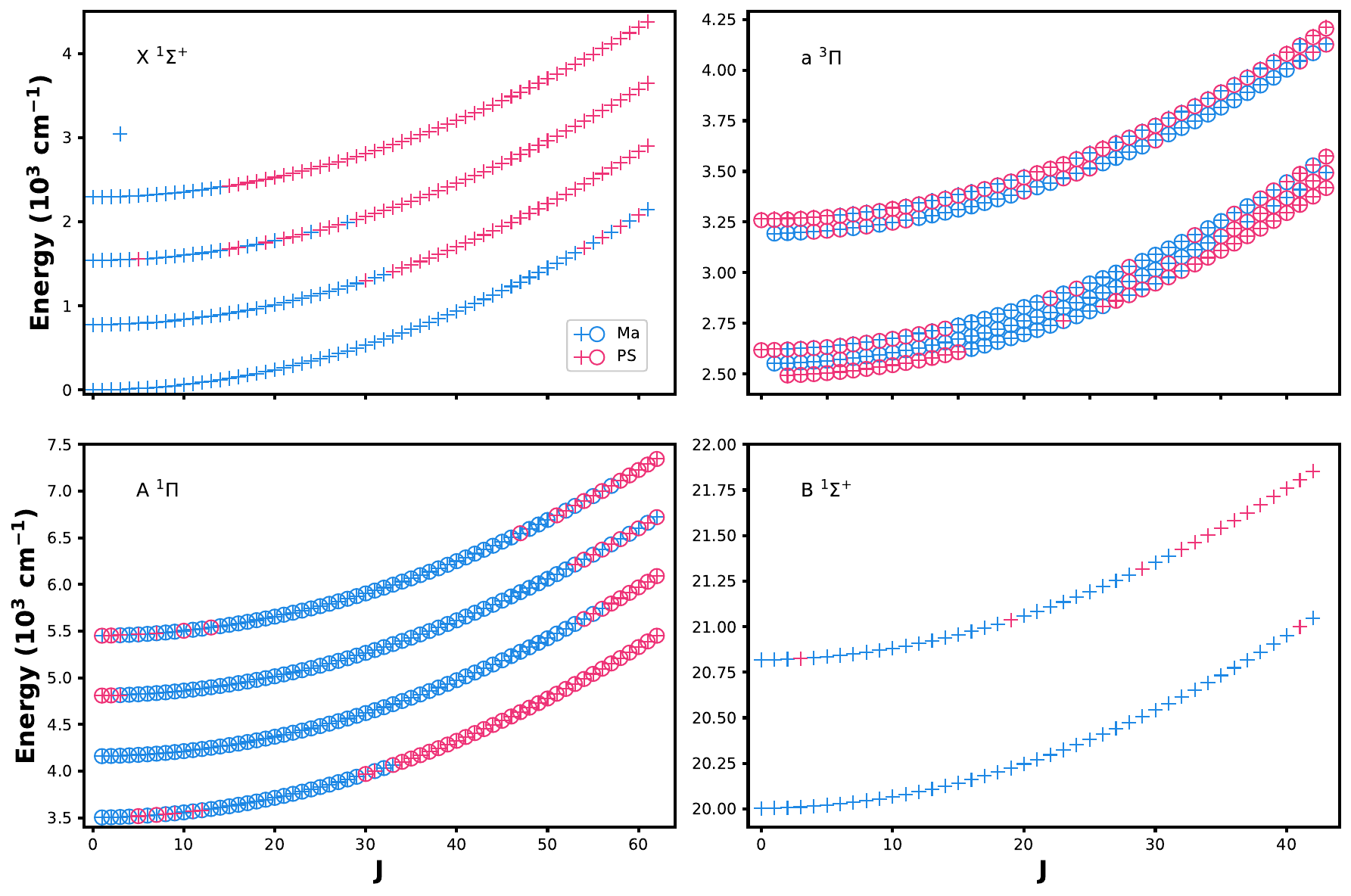}
    \caption{The updated \Marvel{} (Ma) and predicted shift (PS) energies for the \MgX, \MgA, \Mgaa, \MgB{} electronic states of \MgO{} as a function of the $J$ quantum number. The `o' symbol indicates $f$ parity and the `+' symbol indicates $e$ parity.}
    \label{fig:MgO_ELs}
\end{figure*}


\subsubsection{Predicted shift energy levels}

The predicted shift methodology was used to calculate energy levels. Where necessary, uncertainties were extrapolated using $a = 0.0001$~\cm\ (see \cref{e:ps_unc}) and a maximum cutoff for $\Delta\tilde{E}_{\text{PS}}$ of 10 \cm{}.

The way in which the predicted shift levels interpolate and extrapolated from the \Marvel{} energies  to the different rotational levels of the same vibronic state is visualised most clearly in \cref{fig:MgO_ELs}. 

\subsubsection{Isotopologue-extrapolation energy levels}\label{s:isoextmgo}

The line lists for the \mgos, \mgoe, \mgof\, and \mgoff{} isotopologues were updated using the isotopologue extrapolation (IE) correction methodology.
For the energy levels of the main isotopologue which were updated either with a \Marvel{} energy (Ma) or with the predicted shifts (PS), we updated the respective level of the other four isotopologues applying the same final pseudo-experimental correction according to \cref{e:iso}.

In total, 7017 energy levels were updated for each of the five isotopologues for the four lowest electronic states using the IE method. 
The energy uncertainties for the four isotopologues were estimated as twice the uncertainty of the main isotopologue. For the cases without matches to the main (labelled as Ca in the \texttt{.states} file), \cref{eq:unc_ca} was used with a maximum cutoff of 10 \cm, and with $a = 0.0001$ \cm\, $b = 0.01$ \cm\, for all electronic states, and c as the average shift for each electronic state.

For MgO, we are able to directly verify the quality of our isotopologue-extrapolation correction by comparing against experimental data. Specifically, we assessed a small part of the updated line lists for the isotopologues \mgof\ and \mgoff\ for which we found experimental data recorded by \citet{94KaHiTa.MgO} and \citet{86ToHoxx.MgO}:
\citet{94KaHiTa.MgO} recorded transitions in the \MgA\ -- \MgX\ electronic band and the $v_{\rm A}-v_{\rm X} = 1-0$ vibrational band with $J=6-41$ for \mgof\ and $J=6-42$ for \mgoff, and \citet{86ToHoxx.MgO} published six microwave transitions with $J=0-7$ for \mgoff\ with $v=0$.

For \mgoff\, the average residues in positions against experiment for four rotational \MgX-\MgX{} transitions are 1.2 $\times 10^{-4}$ \cm{} without IE correction and 6.7$\times 10^{-6}$ \cm{} with IE correction, while for the rovibronic \MgA-\MgX{} band, the IE correction had a  more modest reduction in errors from 0.027 \cm{} to 0.021 \cm{}. The experimental data mostly agreed with our line list energies within their mutual estimated uncertainties. 


Similar comparisons for \mgof{} rovibronic \MgA\ --\MgX\ transitions found that for the P and Q branches, applying the IE correction led to a reduction in average residual from 0.017 \cm{} to 0.012 \cm{}. However, the residuals associated with the R branch transition were notably larger (0.05 \cm{} average) and did not significantly change with applying the IE correction; the likely cause of this difference is incorrect assignments of the experimental data.

\begin{figure*}
    \subfloat[The energy distribution for the rovibronic states of \MgO{} as a function of energy source type in each electronic state.]{%
        \includegraphics[width=0.45\linewidth]{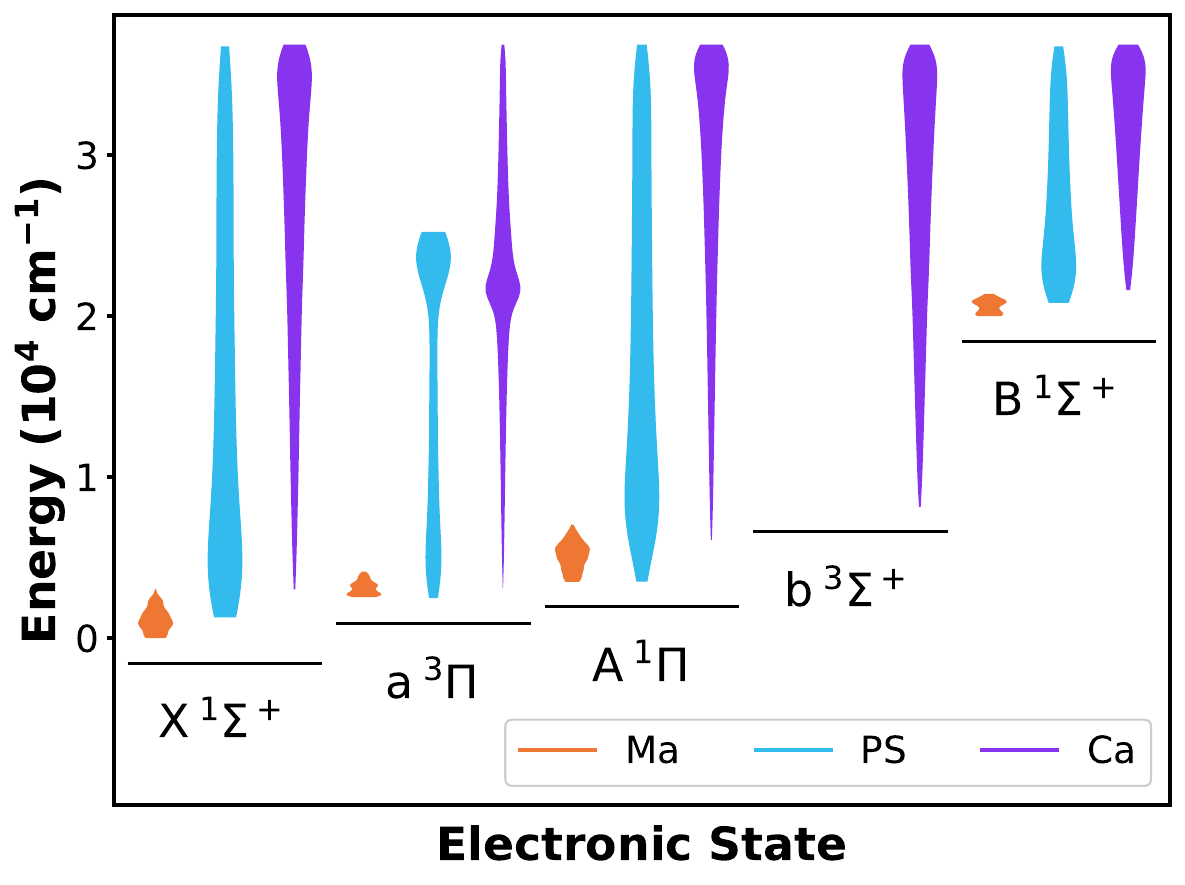}%
        \label{fig:distrib_mgo}%
    }\hfill
    \subfloat[The transition source type for \MgO, i.e. the cumulative density of transitions as a function of intensity depending on energy source type, computed at 2000 K using the program \textsc{ExoCross} \citep{jt708}.]{%
        \includegraphics[width=0.5\linewidth]{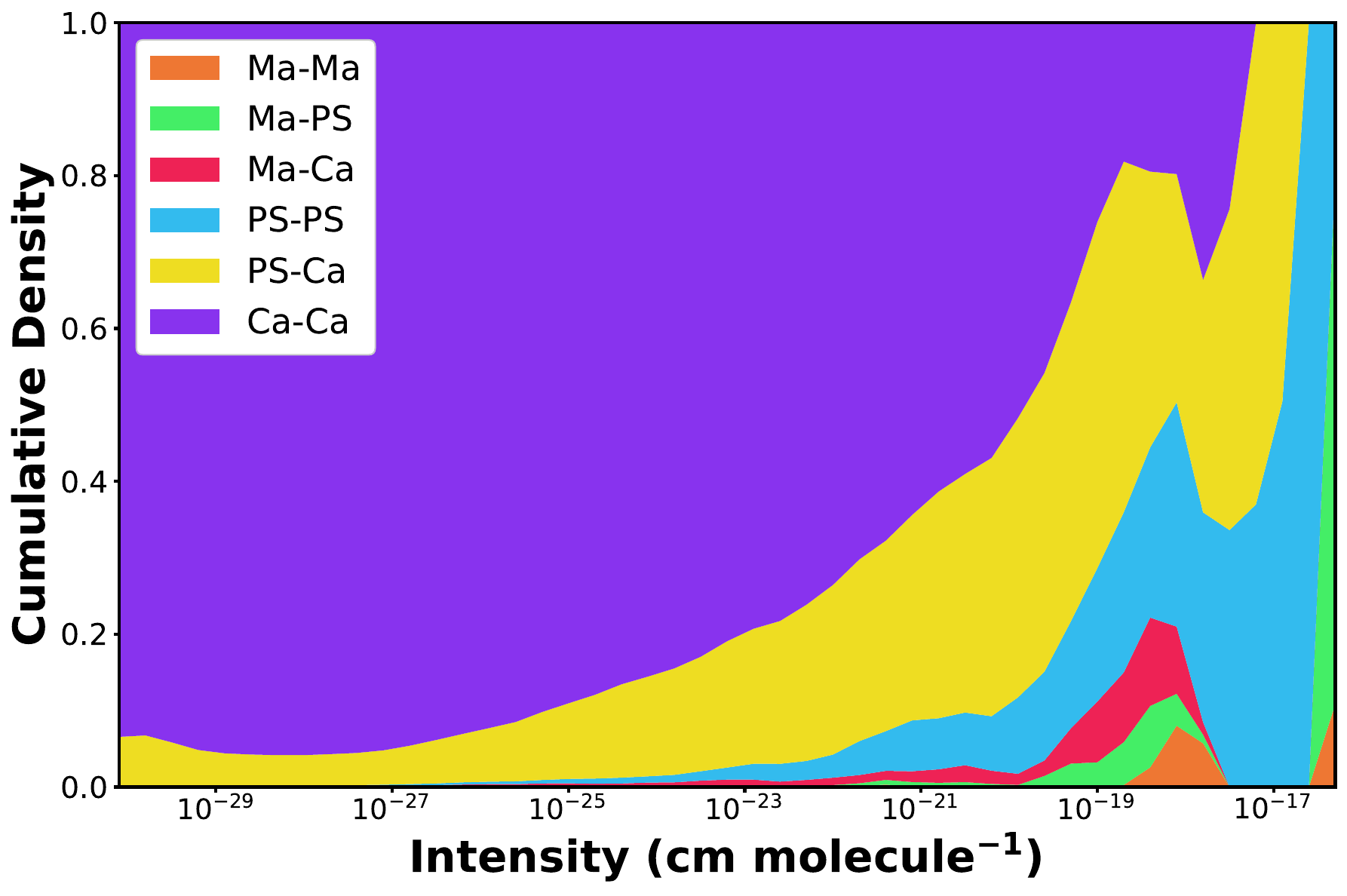}%
        \label{fig:mgo_cumulative}%
    }
    \vspace{1em}

    \subfloat[The \mgo{} absorption cross section computed at 2000 K using the program \textsc{ExoCross} \citep{jt708} with Gaussian line profiles of 1.0 \cm{} half-width half-maximum. The black cross section shows all transitions in the line list with decomposition into dominant electronic bands.]{%
        \begin{minipage}{\linewidth}
            \centering
                    \includegraphics[width=\linewidth]{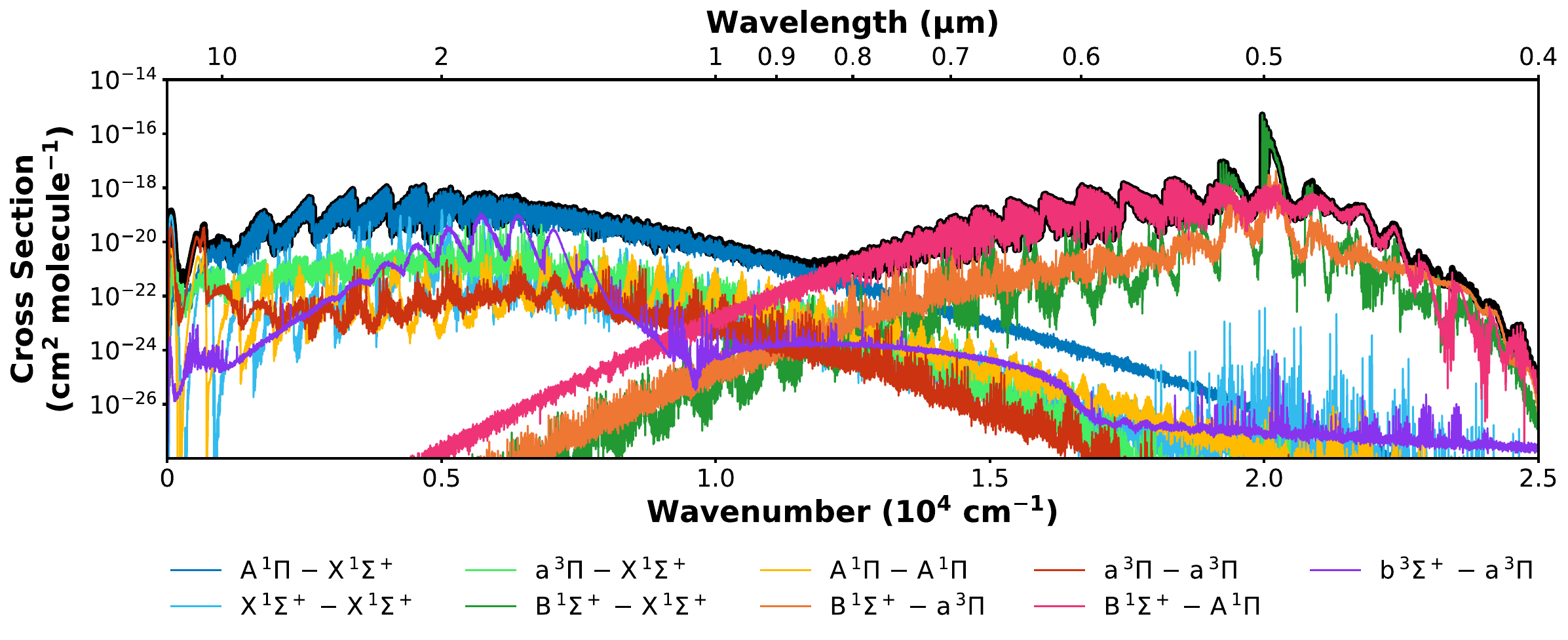}
            
            \label{fig:mgo_decomp_elec}
        \end{minipage}}
        
    \subfloat[
    The black cross section shows all transitions in the line list, whereas the orange cross section shows only \Marvel{} (Ma) experimental transitions (with variational intensities), and the blue cross section shows all possible transitions between \Marvel{} (Ma) and predicted shift (PS) energy levels (with variational intensities).]{%
        \includegraphics[width=\linewidth]{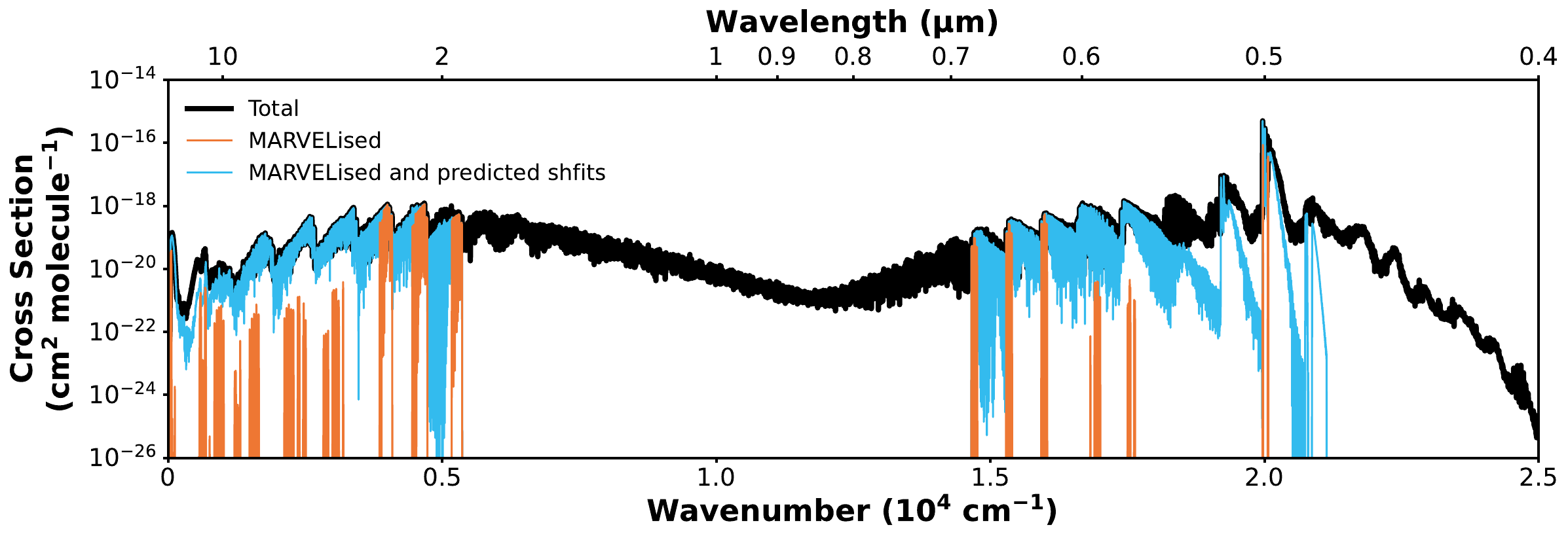}%
        \label{fig:mgo_decomp_ma_ps}%
    }

    \caption{Details of updated LiTY line list for \MgO{}.}
\end{figure*}

\subsection{Line list}

\subsubsection{Construction}

The LiTY line list \citep{jt759} for \MgO{} was updated by incorporating 820 \Marvel{} (Ma),  6197 predicted shift (PS) energy levels into the line list, with 179\,825 variationally calculated (Ca) energy levels retained for vibronic levels without any experimental data. The distribution of energy levels within each electronic state depending on their source is shown in \cref{fig:distrib_mgo}. The pattern that emerges of \Marvel{} energy levels focused on the low lying rovibronic states in each electronic state complemented by the  predicted shift and calculated levels is common for most molecules.  

The \texttt{.states} files for the \mgos{}, \mgoe{}, \mgof{} and \mgoff{} isotopologues were updated using the isotopologue extrapolation (IE) correction methodology as detailed in \cref{s:isoextmgo}.

The \texttt{.trans} files for all isotopologues were cleaned to remove negative transition frequencies. Specifically, 92, 136, 139, 131, and 127 transitions were removed for the \MgO{} \mgos{}, \mgoe{}, \mgof{} and \mgoff{}  isotopologues, respectively.

The absorption cross section between the several electronic bands computed at $2\,000$ K is illustrated in \cref{fig:mgo_decomp_elec}. The strongest transitions around 500 nm are the \MgB{} - \MgX{} band, but the hot band \MgB{} - \MgA{} transition dominates for most of the visible spectra with the \MgA{} - \MgX{} transition dominating the near IR region. The spin forbidden \MgB{} - \Mgaa{} band displays significant intensity due to \Mgaa{} - \MgA{} spin-orbit coupling that mixes their wavefunctions. The experimentally unobserved band \Mgb{}-\Mgaa{} is confirmed to be less intense than the rovibronic \MgA{} - \MgX{} manifold across the full spectral region.

\subsubsection{Suitability for high-resolution cross-correlation}


The transition source type plot (described in \cref{sec:twofigs}) for MgO is shown in \cref{fig:mgo_cumulative}. This peak in Ma and PS involved transitions near 10$^{-18/19}$ is atypical and indicates that some of the weaker spectral bands have been more heavily studied that the spectral band that is strongest at 2000 K. The \Marvel{} data for this molecule is reasonably modest with the predicted shift data is crucial for dramatically extending the number of transitions whose wavenumbers are known to reasonably high accuracy.

The high-res vs total cross-section of MgO is shown in  \cref{fig:mgo_decomp_ma_ps} at $2\,000$ K. In spectral regions where the \Marvel{} and total cross-sections overlap, then the line list is very suitable for HRCC. In the case of MgO, due to the limited experimental data, this region is very small (a few narrow windows between around 1.9 and 3 \um{}, and between 640 - 690 nm). In the absence of perturbations, predicted shift may be sufficiently reliable for HRCC especially for high signal-to-noise observations; in this case, the suitable spectral windows for HRCC in MgO extend to approximately 580 - 690 nm and 1.9 to 3 \um{}).

While cross-correlation techniques are most sensitive to the strongest features in spectra, good coverage across the wavelength region being targeted for cross-correlation is desirable to reduce noise and obtain an unambiguous detection.

Note that this analysis has been done at 2000 K; cooler temperatures would likely increase the suitability of the line list for HRCC analysis since the experimental data is more complete at lower rotational quantum numbers that would be more populated at lower temperatures.

\subsection{Future work}
\cref{fig:mgo_cumulative,fig:mgo_decomp_ma_ps} show clearly that more high resolution experimental data for the four lowest electronic states of MgO are high priority to enable more robust searches for this molecule using high-resolution cross-correlation techniques. In particular, the experimental data was limited to modest rotational quantum numbers, meaning that the band peak and full range couldn't be predicted for the hot 2000 K environments. Nevertheless, the PS methodology allowed extrapolation to higher $J$ than the experimental data, and may be sufficiently reliable to enable HRCC detections of MgO. 

Predicted shift methodology is only feasible if some rotational states of a particular vibronic level are known. Data for the $v=2,3$ \MgB{} state and any levels of \Mgb{} electronic states would be welcome.


\section{\TiOtitle{}: Details and updates are critical}\label{sct:TiO}

\subsection{Background}

Diatomic titanium oxide (TiO) has long been known to be a critical molecular absorber in cool stellar atmospheres with temperatures around 1700 - 2500 K. After initial controversial detections \citep{17NuKaHa.TiO,17SeBoMa.TiO} and some unexpected non-detections \citep{20MeGiNu.TiO}, it is now generally accepted to also be present in at least some hot Jupiter exoplanets including WASP-189 b \citep{22BiHoKi.exo} with further tentative detections in WASP-69 b \citep{23OuWaZh.TiO} and potentially WASP-33 b \citep{21SeNuSt.TiO}.  The presence of TiO in these planets' atmospheres is thought to contribute to the formation of temperature inversions \citep{20PiMaMc.TiO}, where the upper atmosphere is hotter than the lower atmosphere. To understand these complex atmospheric processes, reliable measurements of TiO abundance are required. Given there has been significant controversy surrounding TiO presence in some hot Jupiter exoplanets \citep{21SeNuSt.TiO}, verifying the completeness and accuracy of the TiO spectral data is crucial to enable robust examination of modelling results. New experimental data filling existing gaps should be incorporated into the line list as quickly as feasible \citep{21McKemmishDiatomics}. 

The ExoMol TiO line list, Toto, was published by \cite{jt760} and contained 301\,245 Ca (variational \textsc{Duo}) and 17\,802 Ma (experimentally-derived \Marvel{}) energy levels. In mid-2021, the Toto line list for TiO was updated to correct an oversight in the original 2019 release in which the a~$^1\Delta$ and e~$^1\Sigma^+$ electronic states were initially not \Marvelised{}. This 2021 update replaced the original \Duo{} variational energy levels with experimentally-derived \Marvel{} energy levels for 660 energy levels in the a~$^1\Delta$ and 98 energy levels of the e~$^1\Sigma^+$ state. This update improved the high-resolution accuracy of the line list, particularly in the spectral regions dominated by transitions involving these states, particularly 2000 -- 2500 \cm{} (4 -- 5~\um{}), 3000 -- 3500 \cm{} (2.8 -- 3.3 \um{}), 9000 -- 9500 \cm{} (1.05 -- 1.10 \um{}), 10\,000 -- 10\,500 \cm{} (0.95 -- 1.00 \um{}), 11\,000 -- 11\,500 \cm{} (0.87 -- 0.91~\um{}). 

Here, we provide significant updates to these original line lists for all isotopologues. With the release of this paper, the 2021 update will be superseded by the 2024 update of the Toto TiO line list. All the advantages of the 2021 update will be retained in this 2024 update. 

Specifically, in this 2024 TiO update, we construct a hybridised line list by combining the calculated variational \Duo{} energy levels (Ca) from the 2019 Toto spectroscopic model \citep{jt760}, an updated \Marvel{} compilation (Ma) and new predicted shift energy levels (PS).
This update also provides uncertainty estimates for all energies, which were not present in the original 2019 line list.
The spectroscopic model for TiO has not been refit for this update and thus the \texttt{.trans} file is unchanged from the 2019 or 2021 line lists. As new experimental data becomes available, this is certainly a worthwhile exercise though beyond the scope of this paper. 


\subsection{Data sources for energy levels}

\subsubsection{The 2024 \Marvel{} update} 

A \Marvel{} compilation of experimental transition frequencies into a self-consistent spectroscopic network is an ongoing process that should continue in line with the publication of new assigned experimental data \citep{21McKemmishDiatomics}. For \TiO{}, the \Marvel{} energy levels were first compiled in \cite{jt672} and updated for \cite{jt760} (as part of the 2019 Toto line list construction). 

Herein, we use {\sc Marvel4.1} to expand the TiO \Marvel{} dataset to incorporate new experimental data for the E~$^3\Pi$ -- X~$^3\Delta$ \citep{20BeCaxx.TiO} (659 transitions, all validated) and B~$^3\Pi$ -- X~$^3\Delta$ bands \citep{22CaBe.TiO} (5506 transitions, all validated), as well as new experimental transitions for the X~$^3\Delta$ rovibrational band were included from 19BrWaTu \citep{19BrWaFu.TiO} (16  transitions, all validated) and 21WiBrDo \citep{21WiBrDo.TiO} (514 transitions, all validated).

There were some validation issues; in these cases, the newer data was preferred, and inconsistent older data removed from the final spectroscopic network (most notably, 179 lines from the 73Phillips \citep{73Phillips.TiO} data were invalidated).

\begin{table*}
    \caption{An overview of the 2024 TiO \Marvel{} energy levels (EL) and comparison against \Duo{} variational calculated energy levels (|Ma-Ca|) in \cm. It should be noted that parity is not explicitly considered in the TiO \Marvel{} compilation for $\Delta$ and $\Phi$ states and is instead added when hybridising the full line list.}
    \label{tab:TiOMarvel}
    \begin{tabular}{lcccccccccccc}
    \toprule
    State & $v$ range & $J$ range & \#ELs & Unc. range & Avg. Unc. &  EL range & |Ma-Ca| range & Mean |Ma-Ca| \\ 
    \midrule
    X~$^{3}\Delta$ & 0 - 5 & 1 - 162 & 5000 & 0.0000 - 7.92 & 0.40 & 0 - 14878 & 0.0000 - 5.50 & 0.19 \\
    a~$^{1}\Delta$ & 0 - 3 & 2 - 100 & 638 & 0.0215 - 2.02 & 0.12 & 3446 - 8978 & 0.0001 - 7.45 & 0.12 \\
    d~$^{1}\Sigma^+$ & 0 - 5 & 0 - 92 & 402 & 0.0273 - 0.36 & 0.06 & 5661 - 12259 & 0.0002 - 0.28 & 0.05 \\
    E~$^{3}\Pi$ & 0 - 1 & 0 - 61 & 386 & 0.0100 - 0.10 & 0.04 & 11838 - 13969 & 0.0022 - 1.82 & 0.69 \\
    A~$^{3}\Phi$ & 0 - 5 & 2 - 163 & 5152 & 0.0002 - 8.06 & 0.48 & 14021 - 28825 & 0.0000 - 3.19 & 0.36 \\
    b~$^{1}\Pi$ & 0 - 4 & 1 - 100 & 806 & 0.0245 - 1.20 & 0.08 & 14717 - 20507 & 0.0002 - 1.05 & 0.15 \\
    B~$^{3}\Pi$ & 0 - 2 & 0 - 148 & 2275 & 0.0080 - 3.62 & 0.16 & 16224 - 27727 & 0.0011 - 68.2 & 4.25 \\
    C~$^{3}\Delta$ & 0 - 7 & 1 - 158 & 4756 & 0.0010 - 6.32 & 0.28 & 19341 - 31462 & 0.0001 - 11.1 & 0.50 \\
    c~$^{1}\Phi$ & 0 - 3 & 3 - 101 & 608 & 0.0245 - 2.31 & 0.13 & 21290 - 26685 & 0.0001 - 8.27 & 0.20 \\
    f~$^{1}\Delta$ & 0 - 2 & 2 - 72 & 302 & 0.0245 - 0.15 & 0.05 & 22515 - 25321 & 0.0004 - 0.63 & 0.08 \\
    e~$^{1}\Sigma^+$ & 0 - 1 & 1 - 59 & 98 & 0.0582 - 0.11 & 0.07 & 29960 - 32515 & 0.0107 - 3.57 & 0.77 \\
    \bottomrule
    \end{tabular}
\end{table*}



The 2024 TiO \Marvel{} compilation contains 12\,164 energy levels from 61\,509 validated (62\,935 total) transitions compared to 2019 compilation of 10\,761 energy levels from 51\,547 validated (56\,240 total) transitions.
The scope of the data is summarised in \cref{tab:TiOMarvel}. The most significant difference from the 2019 \Marvel{} dataset is that B~$^3\Pi$, $v$=2 lines are available for the first time; there has also been a significant increase in the $J$ range of data for the E~$^3\Pi$ and B~$^3\Pi$ states as expected. Furthermore, 515 out of 517 lines by \citet{02KoHaMu.TiO} and \citet{91SiHaxx.TiO} in the band E~$^3\Pi$ -- X~$^3\Delta$ with unresolved upper state parities were assigned e and f parities using the \Marvel{} energy levels. No significant changes in the X~$^3\Delta$ state were identified.

\begin{figure}
    \centering
    \includegraphics[width=\linewidth]{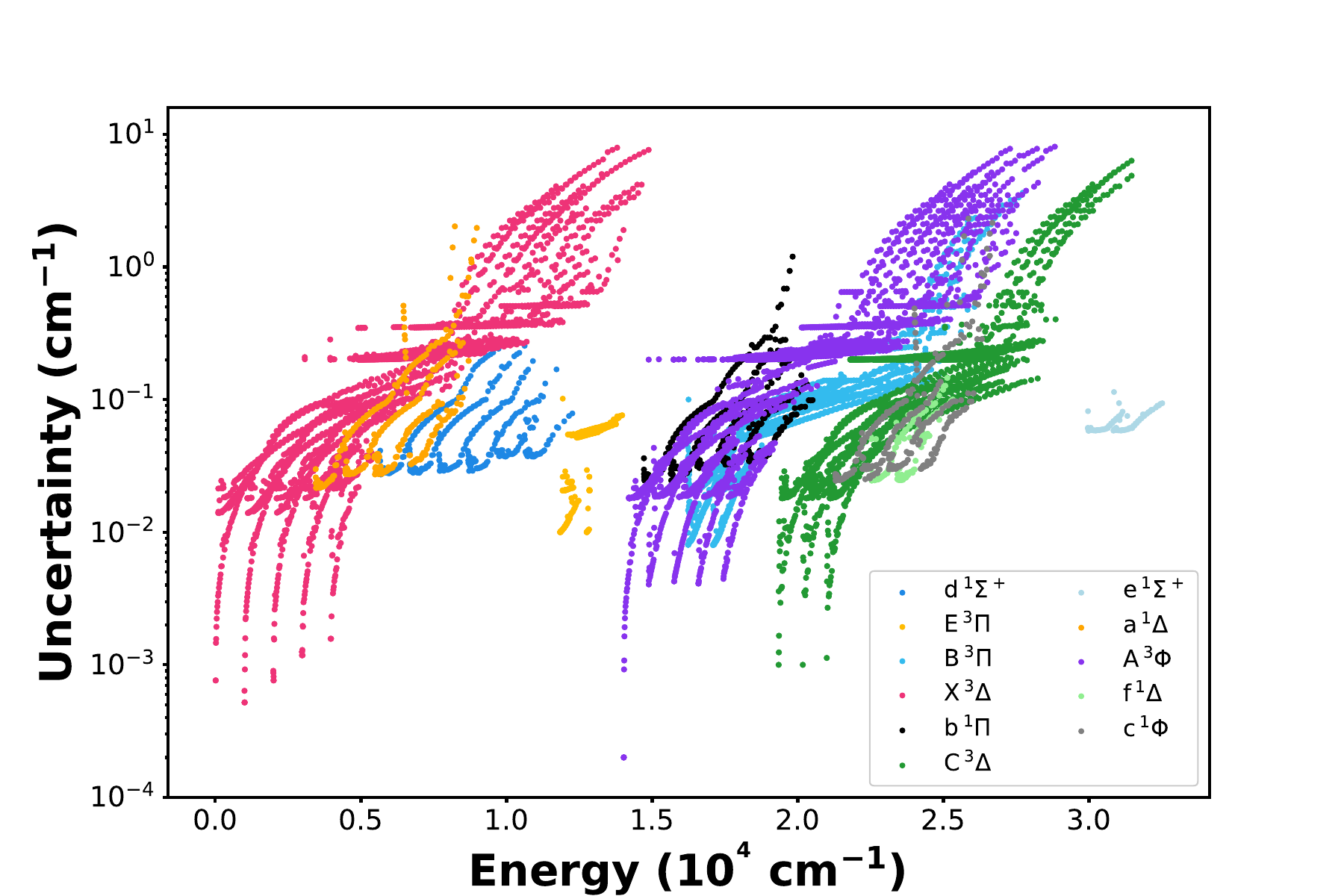}
    \caption{The \Marvel{} (Ma) energy uncertainties for \TiO{} as a function of energy.}
    \label{fig:Tio_unc}
\end{figure}

We use {\sc Marvel4.1} to produce the \Marvel{} energy uncertainties shown in \cref{fig:Tio_unc}.
{\sc Marvel4.1} allows for the uncertainties in the energies to be calculated using a bootstrapping approach, as described by \citet{jt908}.
This method accounts for inconsistencies between multiple transitions to or from a given level by applying an increase to the level's final uncertainty.
This is achieved through randomly increasing the uncertainties of all transitions by a factor between 2 and 10 and assessing the variance in the final energy values obtained over multiple iterations.
Accordingly, the uncertainties in the \Marvel{} energy levels present in the 2024 update that were also present in the 2019 data set have uncertainties on average 14\% larger.
These uncertainty changes have a large standard deviation of 87\% however, with uncertainties in the extreme cases being up to 200 times smaller or 50 times larger in the 2024 \Marvel{} data.
These uncertainties are incorporated into the 2024 TiO line list update for the \Marvelised{} energy levels.

\subsubsection{Calculated variational data}

The spectroscopic model for TiO has not been updated from the 2019 model, so we refer readers to \cite{jt760} for details.
The new \Marvel{} experimental data offers the opportunity to assess the accuracy of the 2019 TiO spectroscopic model for states against which it was not explicitly fit, offering new insight into the quality of the spectroscopic model's extrapolation. Most notably, we have 2394 new \Marvel{} energy levels, including 216 in the E state (in $v=0$) and 1783 in the B state (477, 607 and 699 in $v$=0, 1, 2, respectively). We note that there are small differences in other electronic states as a result of the \Marvel{} procedure and correction of quantum numbers (see below).

The E~$^3\Pi$, $v=0$ state is very well predicted with residuals less than 1.82~\cm{} and uncertainties less than 0.11~\cm{} even up to $J=61$. These errors are well within the anticipated accuracy of the \Duo{} spectroscopic model and are not of concern (indeed this is one of the key reasons for including \Marvelised{} energy levels). While the B~$^3\Pi$, $v=0$ and $1$ states possess comparable residuals for similar $J$, the $v=2$ state was found to diverge from the new \Marvel{} energy levels even for low $J$. The calculated deviation from \Marvel{} energies for $v=2$ is expected as the initial fitting data only included data from $v$ = 0 and 1, and only one fitting parameter was employed in the exponent of the extended Morse oscillator potential. The B state has a maximum residual of 68.19 \cm{} for $J=130$ (the largest residual across the whole line list; 11.08 \cm{} outside the B state). A refit of the full spectroscopic model could be merited based on this new data, but this is left to future work. 

Uncertainties for the calculated energies of all five isotopologues of TiO were estimated using \cref{eq:unc_ca}, taking $a = 0.0001$ \cm{} and $b = 0.05$ \cm{}.
$c$ values were determined for each electronic state and in the primary isotopologue these were based on the mean obs.-calc. value for the \Marvelised{} energies in that state.
For states with no \Marvel{} data, the mean obs.-calc. value for all \Marvel{} data was taken, in an attempt to quantify a global uncertainty for the calculations.
The same approach was used to quantify uncertainties in the four isotopically substituted species of TiO, where the uncertainty estimates were based on the shifts seen in the isotopically extrapolated energies.

\subsubsection{Predicted shift data}
\label{subsec:PSTiO}

Due to the good energy level coverage provided by the \Marvel{} data for \TiO{}, it was found that the use of predicted shifts was well suited for estimating the energy residuals in the small number of missing states in the fine structure components of each vibronic branch. To obtain the extrapolated predicted shift uncertainties from \cref{e:ps_unc}, we employed $a = 0.0001$~\cm, and $\sigma$ values that ranged from 0.0009 cm$^{-1}$ for the X~$^{3}\Delta_{1}$, $v=0$ state to 5.0 cm$^{-1}$ for the A~$^{3}\Phi_{4}$, $v=2$ state.


In this context, the inclusion of predicted shift (PS) energy levels over effective Hamiltonian (EH) data should be evaluated. Notably, \citet{20BeCaxx.TiO} and \citet{22CaBe.TiO} recently produced perturbative line lists for the E~$^3\Pi$ -- X~$^3\Delta$ and B~$^3\Pi$ -- X~$^3\Delta$ bands, respectively, based on an effective Hamiltonian fit of these states. 
Typically, perturbative energy levels (EH) like these are assumed to be less accurate than the experimental \Marvel{} data (Ma), but more accurate than the variational (Ca) data. 

When fit over small gaps in the known energy levels, however, the predicted shift methodology can interpolate with accuracy close to that of the \Marvel{} energies. This is particularly favourable as predictions can be made for perturbed levels without direct consideration of the underlying couplings or resonances, which would need to be explicitly included in the construction of an effective Hamiltonian.
 


For spin vibronic states with experimental data, the PS method (see \cref{subsec:PSTiO}) was found to yield high accuracy with an average uncertainty of 0.5~\cm{} for all $J$ values.

\subsection{Line List}

\subsubsection{Construction}

\begin{figure*}
    \centering
    \subfloat[The energy distribution for the rovibronic states of \TiO{} as a function of energy source type in each electronic state.]{%
        \includegraphics[width=\linewidth]{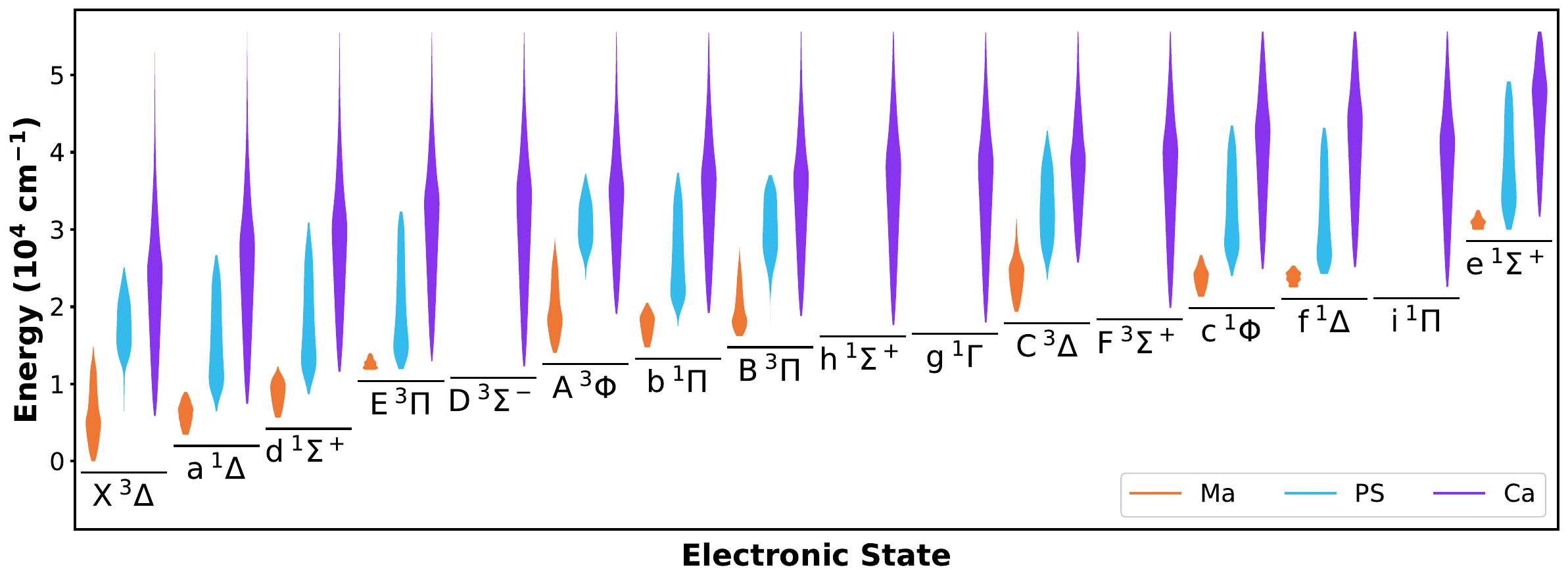}%
        \label{fig:tio_violinstates}%
    }\hfill
    \subfloat[The \TiO{} absorption cross section computed at 2000 K using the program \textsc{ExoCross} \citep{jt708} with Gaussian line profiles of 1.0 \cm{} half-width half-maximum. The black cross section shows all transitions in the new 2024 Toto line list with decomposition into dominant electronic bands.]{%
        \includegraphics[width=\linewidth]{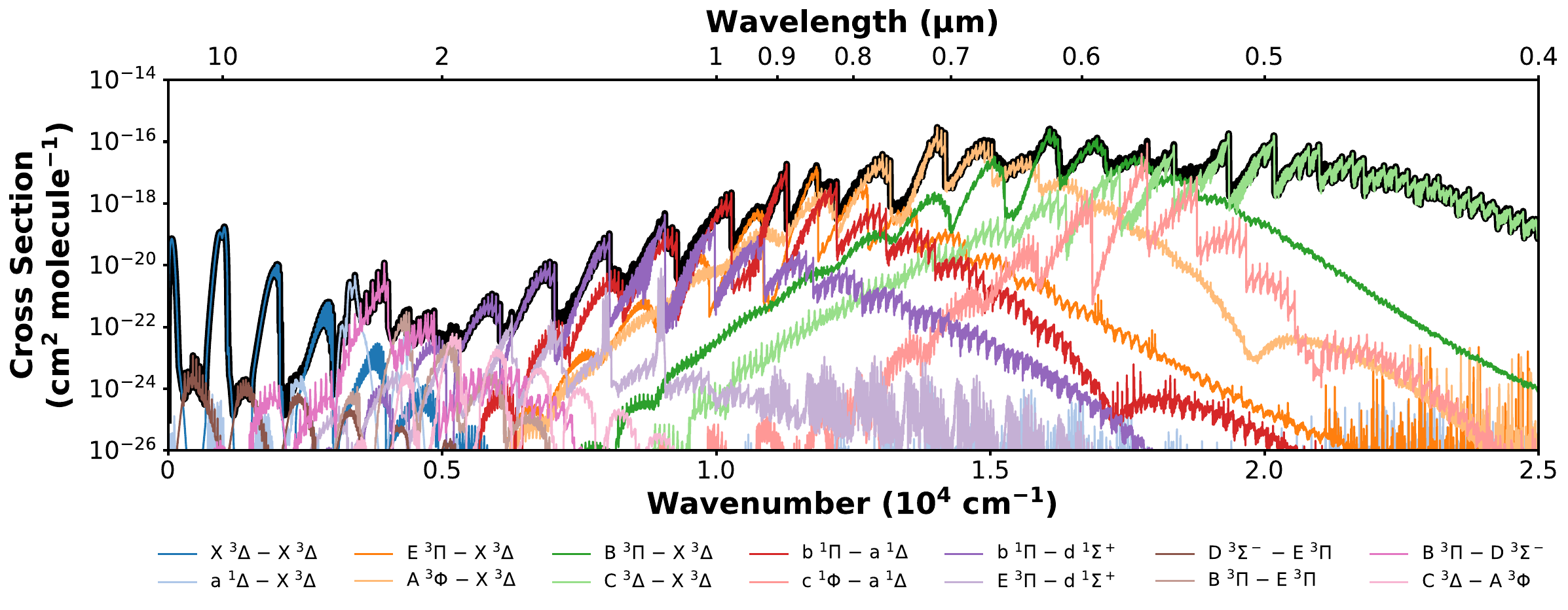}%
        \label{fig:tio_decomp_elec}%
    } \hfill
    \subfloat[
    The black cross section shows all transitions in the new 2024 Toto line list, whereas the pink cross section shows only \Marvel{} (Ma) experimental transitions (with variational intensities). The blue cross section shows the Ma -- Ma cross section produced from the 2021 Toto line list (with variational intensities).]{%
        \includegraphics[width=\linewidth]{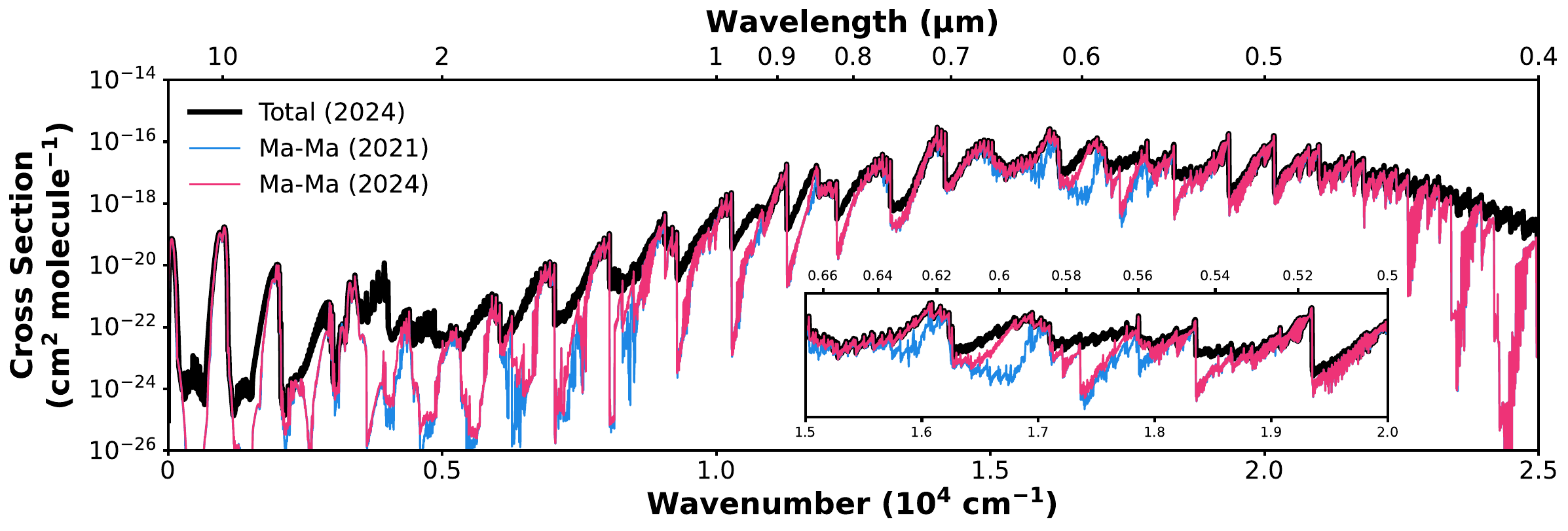}%
        \label{fig:tio_decomp_ma}%
    }\hfill
    \caption{Details of the updated Toto line list for \TiO{}.}
\end{figure*}

The initial step in constructing the hybridised line list was matching the Ca variational energy levels with the Ma experimentally-derived energy levels. A naive energy-hybridisation was extremely problematic, revealing very stark differences in quantum numbers assigned from experiment compared to automatically by \Duo{} with some errors up to 3750 \cm{} due to misassignments. The strategies discussed in \cref{sct:method} were very important in ensuring accurate hybridisation for this molecule. Overall, more than 40\% of rovibronic states have incorrect $v$ and $\Omega$ assignments; most of these are higher energy states and thus unlikely to be \Marvelised{}, but correcting these assignments is also crucial if energy levels extrapolated using the predicted shift or model Hamiltonians are to be used. After application of the predicted shift correction methodology outlined in \cref{sec:PS}, the final energy distribution by source type for each electronic state can be seen in \cref{fig:tio_violinstates}.


The \texttt{.states} files for the \TiOisoa{}, \TiOisob{}, \TiOisoc{} and \TiOisod{} isotopologues were updated using the isotopologue extrapolation (IE) correction methodology. 

The \texttt{.trans} files for all isotopologues were cleaned to remove negative transition frequencies. Specifically, 1147, 1155, 1151, 1690 and 1902 transitions were removed for the \TiOisoa{}, \TiOisob{}, \TiO{}, \TiOisoc{} and \TiOisod{} isotopologues, respectively.

\begin{figure*}
    \centering
    \subfloat[The transition source type plot for \TiO{} using the updated Toto line list, i.e. the cumulative density of transitions as a function of intensity depending on energy source type, computed at 2000 K using the program \textsc{ExoCross} \citep{jt708}.]{%
        \includegraphics[width=0.45\linewidth]{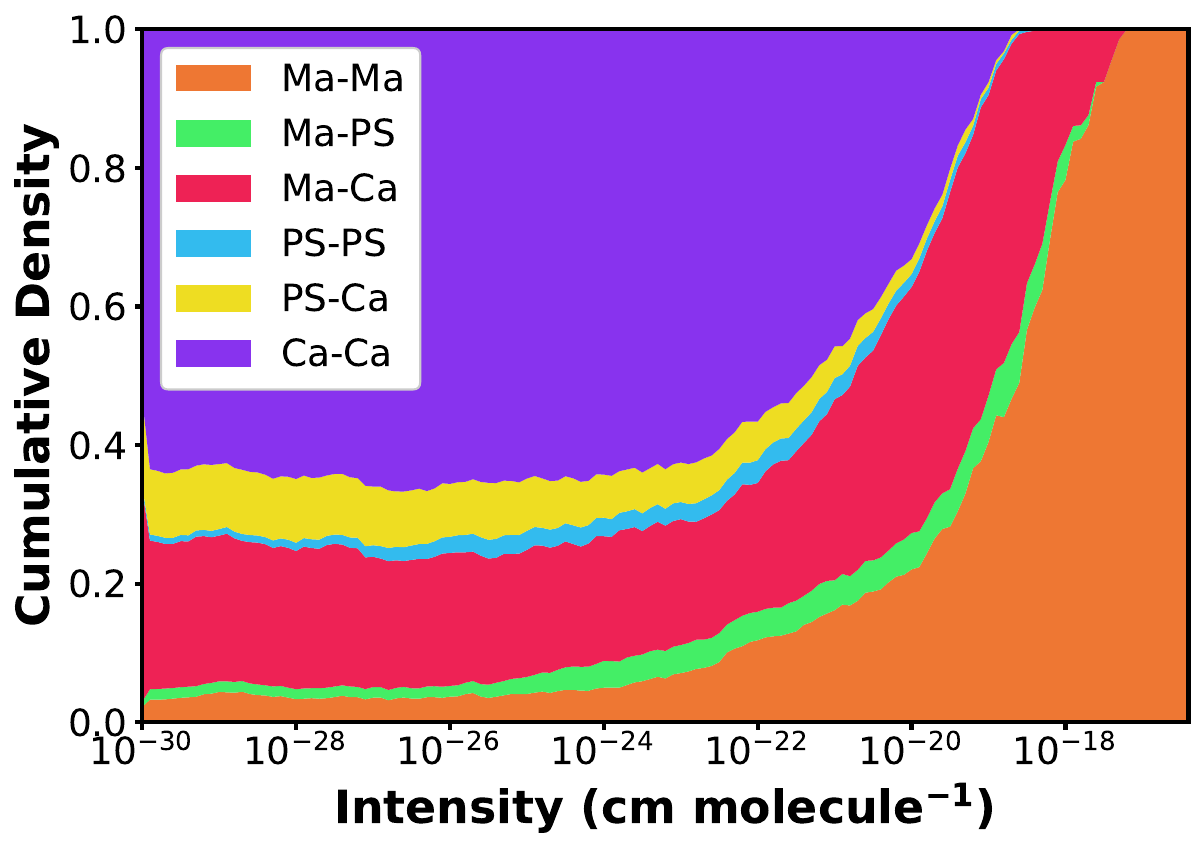}%
        \label{fig:tio_cumulative}%
    }\hfill
    \subfloat[The transition source type plot for \VO{} using the HyVO line list, i.e. the cumulative density of transitions as a function of intensity depending on energy source type, computed at 2000 K using the program \textsc{ExoCross} \citep{jt708}.]{%
        \includegraphics[width=0.45\linewidth]{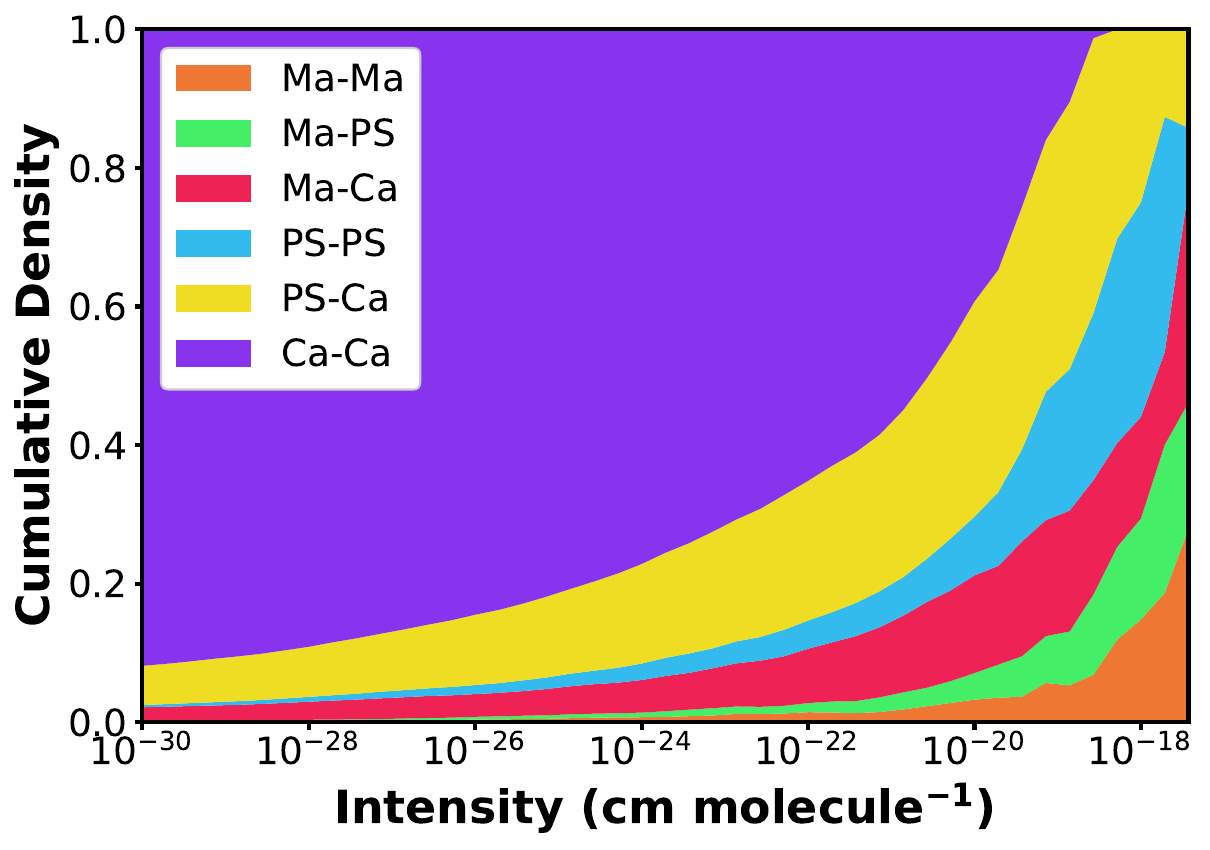}%
        \label{fig:vo_cumulative}%
    }
    \caption{Comparing transition source type plots for \TiO{} from updated Toto and \VO{} from HyVO line list.}
\end{figure*}

\subsubsection{Suitability for high-resolution cross-correlation techniques}

\cref{fig:tio_decomp_ma} shows the high-res vs total cross-section for \TiO{} with the new 2024 Toto line list, as well as the partial cross sections for transitions between only \Marvel{} (Ma) states using the 2024 and 2021 Toto line lists. The \Marvel{} coverage of the strong spectral lines for most of the visible and near-IR region is extremely high for TiO, meaning a wide spectral window is possible for high-resolution cross-correlation (HRCC) studies, enhancing its sensitivity and thus the ability for the line list to detect lower TiO concentrations. This new 2024 update added experimental data supporting HRCC between 590 and 630 nm, which corresponds to the B~$^3\Pi$ -- X~$^3\Delta$ rovibronic transitions.

\cref{fig:tio_cumulative} shows the transition source type plot (described in \cref{sec:twofigs}) as a useful complementary perspective. The diagram confirms that the most intense transitions (above $10^{-18}$ cm/molecule) are more than 90\% determined very precisely by two \Marvel{} experimental levels. However, looking at slightly weaker transitions above 10$^{-19}$ cm molecule$^{-1}$, approximately half of all transitions are between two experimental (Ma) energy levels, where the other half have only one Ma energy level. The width of the red and purple curves demonstrate how important the calculated \Duo{} energy levels are for line list completeness, accurate partition functions and total opacity calculations.The predicted shift energy levels have a small impact but far less than in MgO.

\subsection{Future work}
The very high \Marvel{} coverage for TiO between 450 nm - 1.5 \um{} means that further improvements to this line list are likely low priority compared to other molecules. 

With this new update, there are only very small windows of weaker TiO transitions between 450 nm - 1.5 \um{} where HRCC for TiO at 2000 K may be unreliable; probably most notable is the 565 - 580 nm region which would require experimental data for transitions involving B~$^3\Pi$, $v = 3$ levels. 

Improvements to the Toto spectroscopic model using the new \Marvel{} B~$^3\Pi$, $v=2$  energy levels would improve the quality of the calculated energies for this state significantly by allowing anharmonicity in this state to be modelled based on experimental data for the first time.  This improvement would enhance the accuracy of 565 - 580 nm spectral region, the modelling of some weak transitions and the overall partition function.

\section{\VOtitle{}: Combining \Marvel{} and \Duo{} variational calculated energy levels for complete and accurate line list.}\label{sct:VO}

\subsection{Overview}

Much like TiO, VO is of astrophysical importance in cool stars and hot exoplanets.
It has been observed in the atmospheres of a variety of M-type stars, ranging from subdwarfs \citep{93KiHeLi.VO} and main sequence dwarfs \citep{66SpYoxx.VO,67WiSpKu.VO} to post-main sequence supergiants and hypergiants \citep{77Fawley.VO,71Wallerstein.VO}.
In these atmospheres, the strengths of the VO bands are often used to define the spectral classification of the star \citep{09Bernath.VO}.
In exoplanets like hot and ultra-hot Jupiters, VO is believed to be a strong source of opacity due to the absorption bands in the visible and near-infrared.
This in turn is believed to drive atmospheric thermal inversions and stratospheric heating \citep{09ShFoLi.VO}.

The main isotopologue of VO, \VO{}, is a special case.
As a result of the large nuclear spin, $I = \frac{7}{2}$, of the $^{51}$V atom and the large associated nuclear quadrupole moment, the rovibronic spectrum of \VO{} shows large splittings which cannot be ignored if one wishes to model it for resolving powers greater than roughly $R = 25\,000$.
Through comparisons made by \citet{jt869} between the \VO{} \Marvel{} energies and the calculated \textsc{Duo} energies of \citet{jt644}, it became apparent that simply updating the existing spectroscopic model \citep{jt623} would be insufficient to produce a high-resolution line list.
As is apparent in the unusually large number of hyperfine-resolved transition measurement for \VO{}, an accurate description of the hyperfine effects in the spectra of \VO{} is necessary for a line list intended for use in high-resolution studies \citep{22deKeSn.VO}.
As a result, significant developments have been undertaken to allow the treatment of hyperfine-resolved rovibronic spectra of \VO{}.
These includes allowing for a full (i.e. non-perturbative) treatment of hyperfine effects in \textsc{Duo} \citep{jt855} plus studies on how to best build models which include the explicit treatment of hyperfine effects \citep{jt892}.
Following on from the work of \citet{jt873} to construct a hyperfine-resolved ground state model for VO, and the subsequent 11 electronic state, hyperfine-unresolved model of \citet{jt892}, a new 15 electronic state hyperfine-resolved model was produced \citep{jt912}.
This model refined all of the potential energy curves and the majority of coupling terms present in previous models and crucially introduced a selection of hyperfine coupling terms.

A new, fully hyperfine-resolved line list for VO called HyVO was produced using this new spectroscopic model \citep{jt923}, which is based in part on the updated \Marvel{} networks described below.
The HyVO line list supersedes the widely used VOMYT line list of \citet{jt644} which neglected hyperfine splittings and was not \Marvelised{}.

\subsection{Data sources for energy levels}

\subsubsection{\Marvel{} energy levels}

A \Marvel{} analysis of \VO{} was recently conducted by \citet{jt869}, resulting in two distinct spectroscopic networks as half of the spectroscopic data available was hyperfine resolved.
Accordingly, separate spectroscopic networks were constructed for the hyperfine-resolved and hyperfine-unresolved data sets.
Further, the hyperfine-resolved transitions were deperturbed and added to the hyperfine-unresolved network to supplement its energy level coverage and verify the transitions.
While the hyperfine-resolved network only covered the X~$^4\Sigma^-$, B~$^4\Pi$, C~$^4\Sigma^-$ and 1~$^2\Sigma^+$ electronic states, the hyperfine-unresolved network was additionally connected to the A~$^4\Pi$, A$^{\prime}$~$^4\Phi$, D~$^4\Delta$, 1~$^2\Delta$, 1~$^2\Phi$, 1~$^2\Pi$, 2~$^2\Pi$, 2~$^2\Delta$ and 3~$^2\Delta$ states.
The 1~$^2\Sigma^+$ state is only known via perturbations to the B~$^4\Pi$ state.
The rest of the doublet states were finally connected to the quartet network and hence defined relative to the zero-energy level of the network thanks to the assignment of a spin-forbidden 2~$^2\Pi$--X~$^4\Sigma^-$ band.

A new study \citep{22DoFuGiBr} presented 1\,439 hyperfine-resolved ground state rovibrational transitions in the X~$^4\Sigma^-$--X~$^4\Sigma^-$ (1,0) and (2,1) bands. These transitions were added to the existing set of 6\,643 hyperfine-resolved transitions and passed through \Marvel{} where they were all validated. This new data increases the number of hyperfine-resolved \Marvel{} energy levels from 4\,402 to 5\,702. The primary increase comes from new levels in the previously missing X~$^4\Sigma^-$ $v = 1, 2$ bands, contributing 865 and 413 levels respectively. The extended level coverage also allowed for previously unconnected components of the network to be joined up, adding an additional 2 levels in the B~$^4\Pi$ state and 6 in the C~$^4\Sigma^-$ state.

The hyperfine-resolved transitions of \citet{22DoFuGiBr} were deperturbed into 204 hyperfine-unresolved transitions. The deperturbation procedure has been used during the \Marvelisation{} of the AlO line list \citep{jt835}; the details of the procedure and how the uncertainties in individual hyperfine transitions are propagated are given in \citet{jt869}. The newly deperturbed transitions were added to the existing network of 9\,140 hyperfine-unresolved and deperturbed transitions. All of the new transitions were successfully validated by \Marvel{}. This brings the number of hyperfine-unresolved, empirical energy levels up to 4\,813 from 4\,712.

\subsection{Line List}

\subsubsection{Construction}

The new hyperfine resolved line list for \VO{} was presented by \citet{jt923} and comprises 3\,410\,598 energy levels and 58\,904\,173\,243 transitions.
Of these levels, 28\,760 are \Marvelised{}, 105\,566 were updated based on predicted shifts and the remaining 3\,276\,272 used the term energies calculated by \Duo{}; the distribution of the energy levels as a function of source type can be seen in \cref{fig:vo_violinstates}.



\subsubsection{Suitability for high-resolution cross-correlation techniques}

\begin{figure*}
    \centering
    \subfloat[Energy distribution for the rovibronic states of \VO{} as a function of energy source type in each electronic state.]{%
        \includegraphics[width=\textwidth]{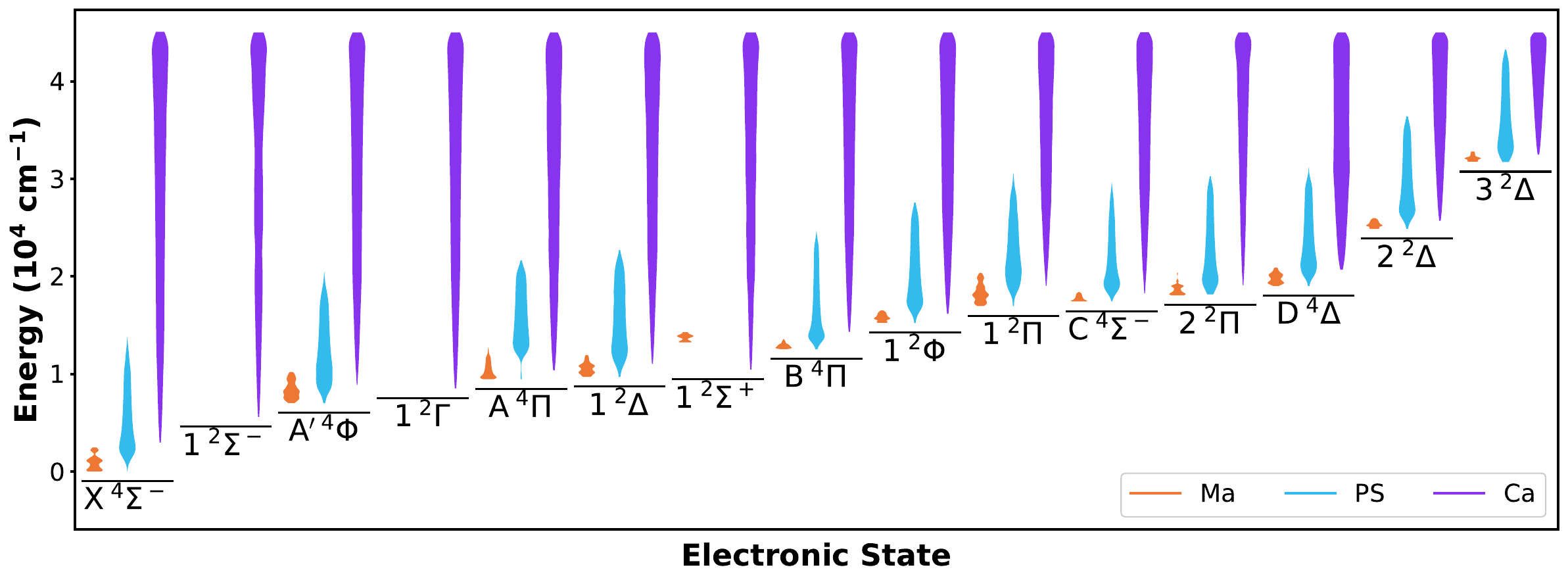}%
        \label{fig:vo_violinstates}%
    } \hfill

    \subfloat[\VO{} absorption cross sections computed at 2000 K using the program \textsc{ExoCross} \citep{jt708} with Gaussian line profiles of 1.0 \cm{} half-width half-maximum. The black cross section shows all transitions in the line list, whereas the orange cross section shows only \Marvel{} (Ma) experimental transitions (with variational intensities), and the blue cross section shows all possible transitions between \Marvel{} (Ma) and predicted shift (PS) energy levels (with variational intensities).]{%
        \includegraphics[width=\textwidth]{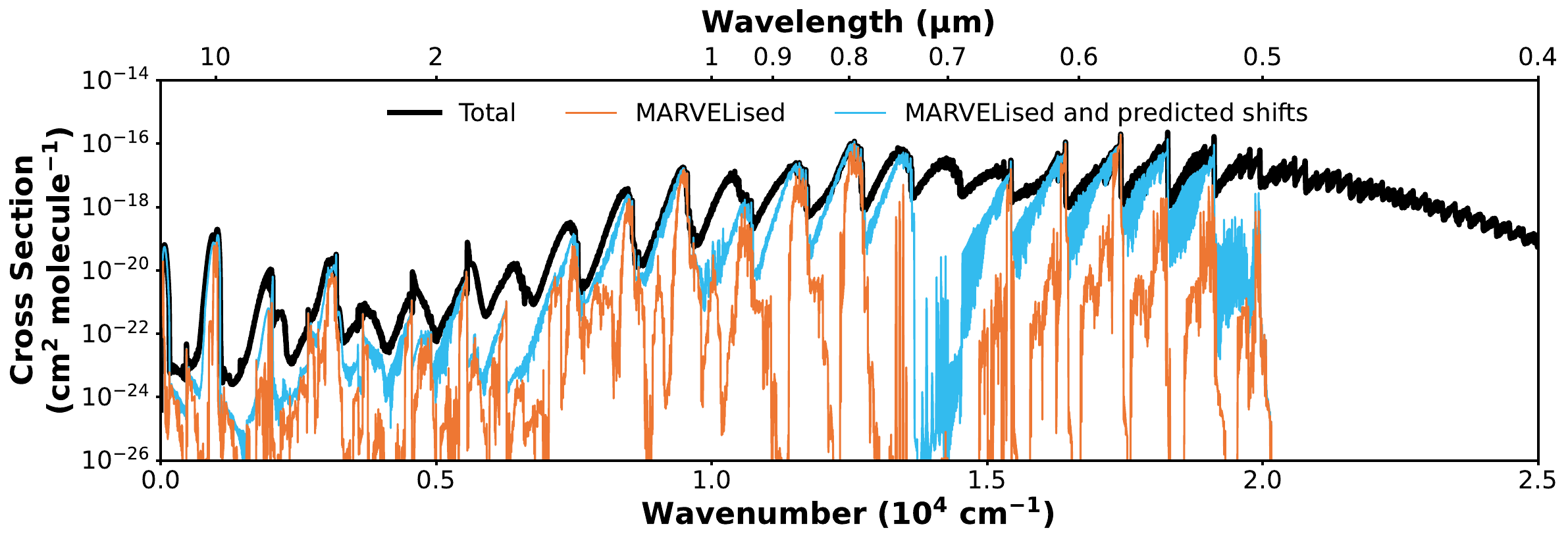}%
        \label{fig:vo_decomp_ma_ps}%
    } \hfill
    \subfloat[Closeup of the visible cross section of \VO{} over the regions recommended for analysis with HRCC techniques, using the HyVO line list. The upper panel shows the contributions from \Marvel{} levels, as well as levels with energies corrected with predicted shifts. The lower panel shows which electronic bands contribute to the total cross section in this region. As with TiO in \cref{fig:tio_decomp_elec}, the total cross section here is comprised of many different electronic bands, with the A~$^4\Pi$--X~$^4\Sigma^-$, B~$^4\Pi$--X~$^4\Sigma^-$ and C~$^4\Sigma^-$--X~$^4\Sigma^-$ bands dominating from roughly 6\,000 \cm{} onwards.]{%
        \includegraphics[width=\textwidth]{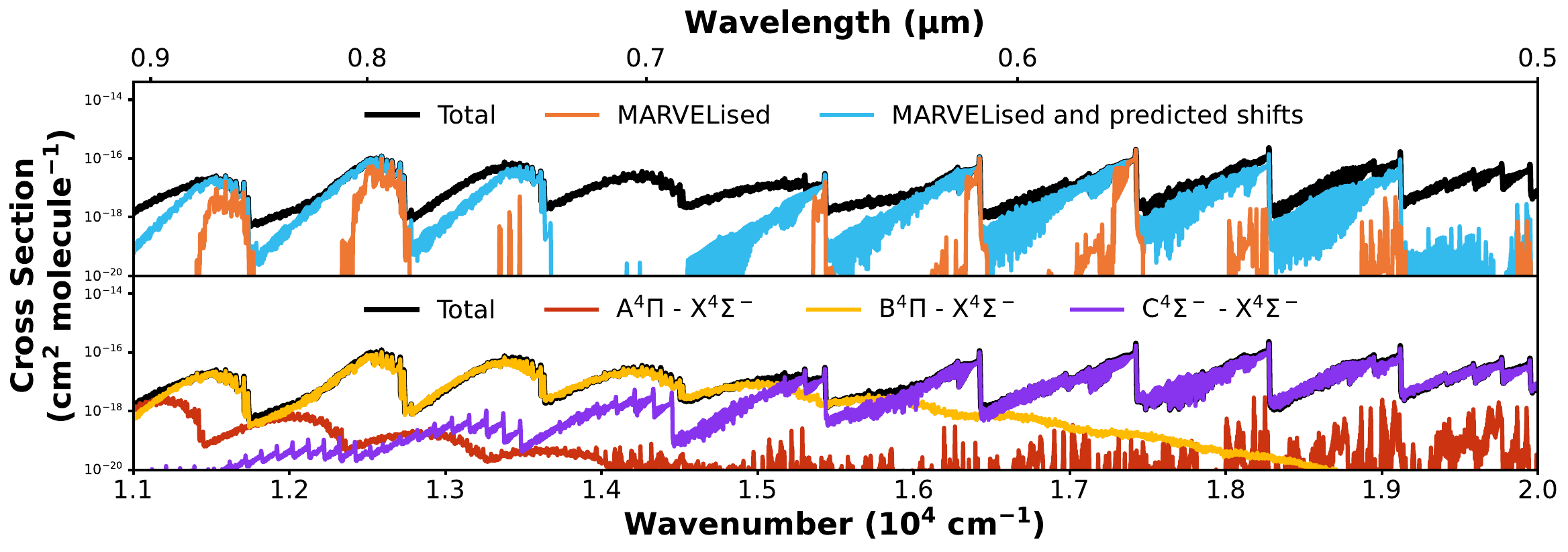}%
        \label{fig:vo_hrcc_decomp}%
    }
    \caption{Details of HyVO line list for \VO{}.}
\end{figure*}

The high-res vs total cross-section shown in \cref{fig:vo_decomp_ma_ps} confirms that the \Marvel{} coverage of HyVO is more extensive than that of \MgO{} shown in \cref{fig:mgo_decomp_ma_ps} but significantly less than the extensive coverage in the updated \TiO{} ToTo line list seen in \cref{fig:tio_decomp_ma}. This conclusion is supported also by the transition source type plot for VO in \cref{fig:vo_cumulative} compared to MgO and TiO. Less than 20 \% of the strong lines (above 10$^{-18}$ cm/molecule for VO have frequencies predicted by transitions between two \Marvelised{} levels. 

The fully \Marvelised{} transitions span a large spectral range and typically cover the strongest part of each spectral band. However, for much of the spectral band, the \Marvelised{} transitions are often an order of magnitude or more less intense than the overall VO cross-section; this might be due to inclusion of only some hyperfine levels. 

Like for MgO, the use of the predicted shift (PS) methodology considerably increases the spectral range for which the cross-section of VO can be used for high-resolution cross-correlation (HRCC) measurements reliably; this can be seen in \cref{fig:vo_cumulative}, \cref{fig:vo_decomp_ma_ps} and \cref{fig:vo_hrcc_decomp}. Transition frequencies obtained by the predicted shift methodology will be of lower accuracy than fully \Marvelised{} transitions. However, for the specific case of VO, many of the predicted shift energy levels were to unmeasured hyperfine levels within the same rovibronic state; the errors associated with this prediction are likely much lower than for other extrapolations as the underlying physics is well understood and predictable.  Conversely, due to the $J$ coverage of the \Marvel{} data, the extrapolation to high $J$ is much larger for VO than TiO, which increases position uncertainty.  

Overall, though some caution is warranted when using predicted shift levels (rather than experimentally-derived \Marvel{} levels) for HRCC studies, it is still of sufficiently high quality that detection using HyVO should be possible for sufficiently high molecular concentration of VO. High-resolution cross-correlation studies targeting the most intense bands, such as those in the visible, are likely to produce the most consistent results. For the case of VO with the HyVO line list, we recommend that astronomers focus on the strongest C~$^4\Sigma^-$--X~$^4\Sigma^-$ bands between 520 - 650 nm or the strongest B~$^4\Pi$--X~$^4\Sigma^-$ between 730 - 900 nm, avoiding the 650 - 730 nm region which contains higher vibrational bands for both electronic transitions unconstrained by \Marvel{} data.

However, since the publication of the HyVO line list, it has been successfully utilised by \citet{24MaGiNu.exo} in a HRCC study of the atmosphere of WASP-76b through transit spectra over the wavelength range 370 - 790 nm.
This study obtained an 8$\sigma$ detection of VO, confirming a previous detection in this ultra-hot Jupiter exoplanet by \citet{23PeBeAl.exo} which utilised the older VOMYT line list, though obtaining stronger signals with the HyVO data.

\subsection{Future work}

The above analysis shows that the suitability of the HyVO line list for HRCC measurements depends on the accuracy of the predicted shift energy levels. Additional experimental data for VO would be highly desirable to reducing this dependence and enhancing the accuracy of the predicted shift levels by reducing the need for extrapolation. We recommend first focusing on extending the coverage of assigned transitions for the X~$^4\Sigma^-$, B~$^4\Pi$ and C~$^4\Sigma^-$ states at hyperfine resolution to high $J$. 

If the predicted shift energy levels prove to be highly reliable in the initial studies and/or the main spectral bands are well characterised, then new measurements should move to characterising higher vibrational levels for these main states in order to broaden the spectral window for VO HRCC measurements, e.g. into the 650 - 730 nm region. These additional measurements of transitions between higher vibrational bands would similarly allow for better constraints on the electronic state potentials and hence more accurate extrapolation to higher vibrational bands, useful for ensuring the accuracy of the partition function and modelling the full cross-section for non-HRCC applications.



\section{Conclusions}

While line list intensities will continue to be produced using variational approaches based on energy and intensity spectroscopic models, the future of line list energies is looking to be far more diverse with many data sources combined to take advantage of their strengths. This updated approach is crucial to meet the current needs of the astronomical community in terms of high completeness and high position accuracy of strong spectral lines, particularly with new telescopes (\eg{} JWST, large ground-based instruments) and new techniques (most notably high-resolution cross-correlation).

Here, we consolidate recent developments in line list construction methodology into a single framework, which we call the hybrid approach. In brief, this approach involves constructing a variational line list based on an energy spectroscopic model (usually \abinitio{} data fitted to experiment) and intensity model (usually purely \abinitio{}) which produces a \texttt{.states} file with rovibronic energies and \texttt{.trans} file with transition intensities. The calculated energy levels within \texttt{.states} file are then updated with other data sources (\eg{} \Marvel{} experiment, model Hamiltonians, HITRAN, \Mollist) or corrections (\eg{} predicted shifts, isotopologue-extrapolation) to produce the best currently known unified source of rovibronic energy levels for that molecule. We select a single terminology and associated acronym for each data source or correction that we apply uniformly across all ExoMol-hosted diatomic line list to this new format to ensure consistency and clear communication. This approach is extensible; new data sources and corrections to energy levels can readily be added. Note that the ExoMol format and this hybrid approach ensure that transition positions arise solely from the energy levels and cannot be updated independently. 

Adding uncertainties for each data source is crucial and indeed is the way that the ``best'' data source is selected for each energy level (this process can sometimes be implicit but making it explicit is helpful). However, this is always, of course, an estimate and there are unexpected challenges that can arise. The most notable example here is when uncertainties of energy levels (\eg{} \Marvel{}'s global uncertainties) substantially overestimate the uncertainty of the observable transition position; manual judgement is currently required here.


We demonstrate this new approach using MgO, TiO and VO as exemplar molecules through analysis of the aforementioned energy source types with careful treatment of uncertainties in all cases. Of particular importance is the suitability of these updated line lists in high-resolution cross-correlation (HRCC) studies, which take advantage of the very high spectral resolving power of ground-based telescopes to make molecular detections despite poor signal-to-noise in observations.

For MgO, we have updated the existing variational line list with experimental \Marvel{} energy levels, and the predicted shifts and isotopologue-extrapolation corrections. While the previous line list was completely unsuitable for HRCC, this new line list can be used with reasonable confidence for HRCC applications in the strong transition around 500 nm and in the broad spectral region 580 -- 680 nm. We note that additional experimental data at higher $J$ and $v$ in measured electronic states would be useful to verify the quality of the predicted shift levels as there is significant extrapolation from experimentally measured data for this molecule. 

For TiO, we have updated the previous \Marvel{} compilation through invalidation of old experimental data and inclusion of new experimental data for the X~$^3\Delta$ rovibrational, B~$^3\Pi$ -- X~$^3\Delta$ and E~$^3\Pi$ -- X~$^3\Delta$ bands. This updated \Marvel{} dataset was then used to rehybridise the 2021 Toto line list for each isotopologue with novel predicted shift calculations. In this case, the high quantity of experimental data meant that predicted shift was only needed to essentially fill gaps in observations, giving it very high reliability. However, to hybridise this line list, a very careful treatment of quantum numbers was required to align the \Duo{}-computed quantum numbers to those derived from energy ordering arguments; intrinsically this reflects the approximate nature of quantum numbers for heavily coupled states but practically this can be a significant challenge when hybridising energy states files. Furthermore, for the B~$^3\Pi$ electronic state, comparison of the original variational spectroscopic model (which was fitted on only $v=0,1$) with the new experimental data (including $v=0,1,2$) did reveal that refitting of the  B~$^3\Pi$ electronic state in the spectroscopic model will be desirable in the future. Regardless, the current 2024 Toto line list has extremely high coverage of experimental data for TiO strong lines across almost all of the visible and near-IR region accessible for ground-based HRCC, with only minor improvements available near 570 nm. Compared to 2021 Toto, this updated dataset provides better experimental coverage in the visible region around 600 nm (17000~\cm{}). 

Importantly, for TiO, the very high quality of the available line list in terms of both completeness and accuracy of strong lines (from experiment) means there should be no concerns about spectral line list quality when analysing JWST observations or conducting HRCC studies; if TiO is present with absorption above the noise limit of the observation, it should be detected with current data in most visible spectral regions (\cref{fig:tio_decomp_ma}). 

For VO, we outline a new \Marvel{} compilation and hyperfine-resolved variational line list. New figures are presented that enable additional insight into the suitability of this line list for HRCC applications. We find that careful selection of spectral region is much more important for VO than TiO as only the very strongest bands around in the near IR and those around 520 - 650 nm are suitable for reliable HRCC studies. The spectral lines with frequencies obtained by predicted shift are also very important for providing a sufficiently large spectral window for effective HRCC measurements. Additional experimental data considering vibronic transitions involving the B~$^{4}\Pi$ state with $v \ge 2$ would be the most beneficial for enhancing the HRCC potential of the VO line list, opening up the spectral region 650 - 730 nm for analysis. Further, experimental assignments of higher $J$ populations for some VO spectral bands would reduce the need to use predicted shifts or validate the accuracy of this approach more closely.  

We note that experimental efforts are beginning to be undertaken which are explicitly designed to improve the quality of the 
MARVEL networks for certain key molecules, thus far for water \citep{20ToFuSiCs} and H$_2$CO \citep{jt906}. This activity, and in particular
extending it to rovibronic spectra of diatomics such as those considered here, would be beneficial for the overall quality of the available
line lists. Finally, the whole ExoMol database is vast, in the region of 10$^{13}$ individual lines \citep{jt939}. A new database, ExoMolHR, has been built \citep{jtHR} which extracts energy levels determined with low uncertainty
($\Delta\tilde{E} \leq 0.01$~\cm) and provides data on lines predicted to high resolution ($R \geq 100~000).$

\section*{Acknowledgements}


This research was undertaken with the assistance of resources from the National Computational Infrastructure (NCI Australia), an NCRIS enabled capability supported by the Australian Government.  The work performed at UCL and the ExoMol project is supported by the European Research Council (ERC) under the European Union's Horizon 2020 research and innovation programme through Advance Grant 883830 (ExoMolHD).

The authors declare no conflicts of interest.

\section*{Supporting Information} 

The \Marvel{} energy, transitions and segment files for \mgo{}, \TiO{} and \VO{} are included as supporting information.

\section*{Data Availability}

The most accurate line list data for the diatomic molecules discussed here will always be available on the ExoMol website (\href{https://exomol.com/}{https://exomol.com/}).

The \Marvel{} energy and transitions files for \mgo{}, \TiO{} and \VO{} are included as supporting information. The \Duo{} spectroscopic models for these molecules are unchanged from previously published data and are available on the ExoMol website and the original publications. 



\end{document}